\documentclass[a4paper,fleqn,usenatbib]{mnras}
\usepackage{amsmath}
\usepackage{txfonts}
\usepackage{caption}
\usepackage[T1]{fontenc}
\usepackage{ae,aecompl}
\hypersetup{colorlinks, citecolor=green, filecolor=black, linkcolor=blue, urlcolor=blue }

\usepackage{graphicx}
\usepackage{amssymb}
\usepackage{dblfloatfix}
\usepackage{breqn}
\usepackage{hyperref}
\hypersetup{
    colorlinks=true,
    linkcolor=blue,
    filecolor=magenta,      
    urlcolor=cyan,
}

\title[The UV/X-ray relation in NGC 4151]{Discarding the disc in a changing-state AGN: the UV/X-ray relation in NGC 4151}

\author[Mahmoud \& Done]{
Ra'ad D. Mahmoud\thanks{E-mail: ra'ad.d.mahmoud@durham.ac.uk} \&
Chris Done
\\
Centre for Extragalactic Astronomy, Department of Physics, University of Durham, South Road, Durham DH1 3LE, UK
}
\date{Accepted XXX. Received YYY; in original form ZZZ}

\pubyear{2019}

\hypersetup{draft}
\begin{document}
\label{firstpage}
\pagerange{\pageref{firstpage}--\pageref{lastpage}}
\maketitle

\begin{abstract}
Recent monitoring campaigns designed to map the accretion regime in AGN show major discrepancies with models where the optical/ultraviolet (UV) is produced by X-ray-illuminated, optically thick disc material within a few hundred gravitational radii. However, these campaigns only monitored X-rays below $10$~keV, whereas the bolometric luminosity for most of these AGN peaks above $50$~keV. We use data from the recent multiwavelength campaign by \cite{E17} on NGC 4151 - the only AGN bright enough to be monitored at higher energies with \textit{Swift} BAT. We develop a spectral-timing model with a hot corona, warm Comptonisation, and outer standard disc. This fits the time-averaged spectrum well, but completely fails to match the UV variability predicted from the X-ray lightcurve. However, it reveals that NGC 4151 had a bolometric luminosity around $1.4\%$ of the Eddington luminosity during this campaign, close to the luminosity at which AGN show a `changing-state' transition, where the broad optical lines disappear. Stellar mass black holes show a state transition at a similarly low Eddington fraction, which is broadly interpreted as the inner disc being replaced by an optically thin flow. We find that the UV lightcurve can instead be matched by reprocessing of the X-ray flux on size scales of the broad line region (BLR; $1.5-20$ light-days) and rule out there being optically thick material inwards of this, as expected if the thin disc is replaced by the flow below the inner radius of the BLR. These results emphasise the need for even longer-timescale, multiwavelength monitoring campaigns on variable AGN.
\end{abstract}



\section{Introduction}
\label{sec:INTRODUCTION}
Active galactic nuclei (AGN) are powered by accretion onto a supermassive black hole, and the energy released makes them the most luminous persistent sources in the sky in the optical, UV and X-ray bands. In the standard picture of accretion, the gravitational energy is dissipated in an optically thick, geometrically thin disc which extends down to the innermost stable circular orbit (ISCO). In this disc, the temperature increases radially inwards, so that the total spectrum is the sum over all radii of the local blackbody spectra produced by each annulus (\citealt{SS73}). The peak disc temperature is linked to the mass of the black hole and its luminosity by $T_{peak}\sim 2\times 10^{5}(\dot{m}/m_8)^{1/4}$~K where $m_8=M_{BH}/10^8M_{\odot}$ and $\dot{m}$ is the mass accretion rate as a fraction of the Eddington rate. These typically far-UV temperatures mean that the bulk of the accretion power in bright AGN ($\dot{m}>0.01$) is hidden by interstellar absorption. In the absence of this absorption, the optical/near UV should be dominated by a sum of blackbodies from the disc, giving a differential flux $F_{\nu} \propto \nu^{1/3}$. However, most optical/UV spectra of AGN are redder than this simple prediction, with mean $F_{\nu} \propto \nu^{-1/2}$ (\citealt{RHV03}). Some of this may be due to circum-AGN dust (\citealt{DWB07}; \citealt{BSP16}), but there is also strong evidence of a marked downturn in the intrinsic far UV spectrum (see e.g. \citealt{ZKT97}; \citealt{TZK02}; \citealt{SSD12}; \citealt{LD14}). 

The X-ray spectra reveal another discrepancy with simple disc models. First, there is an X-ray power law tail which extends out to between a few tens and a few hundred keV. There is a general consensus that the presence of the X-ray tail indicates that some fraction of the gravitational power is released in an optically thin region, distinct from the optically thick disc. Electrons in this `hot corona' can be heated to much higher temperatures than in the dense disc, producing the power law tail by Compton upscattering of seed photons from the disc (as is also seen in the black hole binaries; see e.g. the review by \citealt{DGK07}). However, there is additional soft X-ray emission below $1$~keV: the soft X-ray excess (e.g. \citealt{PRO04}). The origin of this is not well understood, but it seems to connect to the UV downturn, perhaps indicating an additional warm Comptonising region (\citealt{MBR11}, 2015; \citealt{JWD12}; \citealt{POP18}). 

One way to understand this structure is to use variability timescales to map the different regions. The hard X-ray emission varies rapidly with timescales of less than $10-100$~$R_g/c$ (where $R_g=GM_{BH}/c^2$; \citealt{GRV79}; \citealt{FKR}), indicating that the size scale of the hard coronal region is small. Timescales for variability in the standard disc models are very much longer, so any rapid optical/UV variability should be produced by reprocessing of the variable X-ray illumination. Thus tracking the relation of the reprocessed optical/UV emission to the driving X-ray flux variability gives a diagnostic of the structure of the outer accretion disc.

This reverberation technique is well established for deriving the size scale of the broad line region (the BLR; see e.g. the review by \citealt{PH04}), but the recent intensive optical/UV/X-ray monitoring campaigns with the Neil Gehrels Swift Observatory (hereafter \textit{Swift}) have revolutionised the search for \textit{continuum} reverberation from the outer disc. However, the first campaign on NGC 5548 showed the level of disconnect between reprocessing models and observations. Firstly, it showed there was very little correlation between the X-ray variability and the optical/UV variability. In addition, while the UV and optical variability were well correlated with each other, the lag between them was larger than predicted by the disc models by a factor $\sim 2$ (\citealt{E15}).

Subsequent intensive monitoring campaigns on other AGN have only reinforced these two problems, highlighting our lack of understanding of the accretion disc structure rather than revealing it (\citealt{A09}; \citealt{MBR11}; \citealt{L18}). The X-ray variability is generally poorly correlated with the optical and UV variability, in conflict with the idea that the optical and UV are produced by reprocessed X-ray illumination (\citealt{E17}, hereafter E17; \citealt{BLA18}), and while the optical is well correlated with the UV, the lag between them is larger than predicted by the standard disc models (E17).

However there is much more information in the monitoring campaign data than just the lags between the variability seen in different wavebands. Incorporating the spectral information is important, as it constrains the power in the X-rays relative to the optical/UV. Given an assumed disc geometry, this also predicts how much X-ray luminosity intercepts the disc, at which point one can model the disc reverberation. \citealt{GD17} (hereafter GD17) developed the first full spectral-timing model and used it to predict the reprocessed variability from the observed X-ray variability in the \cite{E15} \textit{Swift} NGC 5548 dataset. These predictions completely fail to reproduce the observed optical and UV lightcurves. Even allowing the disc size scale to vary does not help.

GD17 proposed two possible solutions for the first problem, that the observed X-ray variability did not look like the observed UV variability. One was that the model is fundamentally incorrect, and the hard X-rays do not illuminate the optical/UV disc, possibly due to vertical structure on the disc which shields the outer disc from direct irradiation. The second possibility was that the model is essentially correct in assuming that the optical and UV variability is driven by reprocessing of the variable hard X-ray emission, but that the observed \textit{Swift} X-ray Telescope (XRT) X-ray lightcurve did not track the true bolometric X-ray flux variations. This is clearly possible in NGC 5548 as the bolometric flux peaks at $100$~keV (\citealt{MBR11}), whereas the \textit{Swift} XRT band extends only up to $\sim 10$~keV. However, the longer than expected lags between the optical and UV then also required that the material being illuminated has a different size and geometry than expected from an outer disc.

\begin{figure}
	\includegraphics[width=\columnwidth]{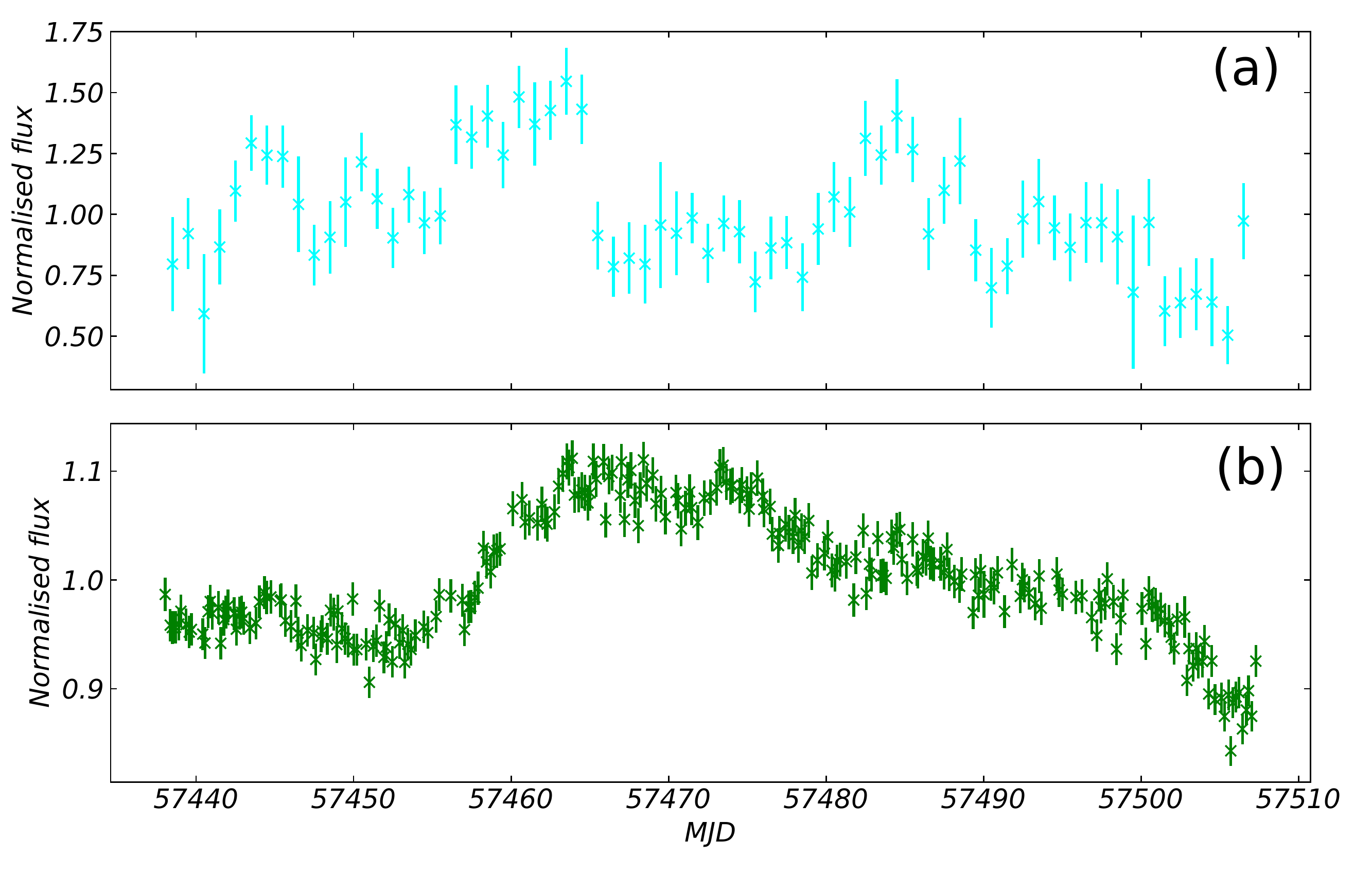}
	\caption{Mean-normalised light curves from the NGC 4151 \textit{Swift} campaign in \textit{Swift} BAT (panel~a) and \textit{Swift} UVOT UVW1 (panel~b; E17).}
	\label{fig:data_curves}
\end{figure}

Here, we apply the procedure of GD17 to the intensive monitoring campaign data from one of the brightest hard X-ray AGN, NGC 4151 (E17). This is the only AGN where the higher energy \textit{Swift} Burst Alert Telescope (BAT) can get reasonable constraints at $15-50$~keV, simultaneously with its optical/UV from the \textit{Swift} UV/Optical Telescope (UVOT) and the $0.5-10$~keV from the XRT. This allows us to track most of the bolometric luminosity directly, so we can for the first time test the fundamental assumption of reprocessing. These data span 69 days with sampling of $~6$~hrs in the UVOT and $\sim 2$~days in the BAT; lightcurves for both the BAT and the representative UVW1 band are shown in Fig.~\ref{fig:data_curves}.

In Section~\ref{sec:modeling} we model the mean broadband spectrum. This is dominated by the hard X-ray luminosity rather than a UV disc component and the inferred bolometric luminosity is below $0.02~L_{Edd}$. Stellar mass black hole binaries are observed to undergo a dramatic transition at this luminosity, from disc-dominated to Compton-dominated spectra. This transition is generally interpreted as the inner thin disc evaporating into an optically thin hot flow (e.g. \citealt{DGK07}). While the bright AGN spectra do not match as well to disc models as the stellar mass black hole binaries, they do appear to make a similar transition to Compton-dominated spectra at a similar luminosity, with the soft X-ray excess collapsing, correlated with the loss of the characteristic broad optical lines (changing state AGN: \citealt{ND18}; \citealt{RAE19}). It seems most likely that NGC 4151 was close to a changing state event at the time of the monitoring campaign, as has been seen in the past from this source (\citealt{PP84}; \citealt{SPC08}).

Nonetheless, there is still optical/UV emission from the AGN, which we model with an outer disc/warm Comptonisation region. A strong UV contribution should be produced at radii of $100-400~R_g$ in this structure, i.e. at a light travel time of less than one day from the the hot inner flow. We predict the UV lightcurve resulting from illumination of this structure by the observed variable hard X-ray lightcurve from the \textit{Swift} BAT, and compare this to the observed UVOT lightcurve both directly and via cross-correlation (Section \ref{sec:modeling2}). As in NGC 5548, the match is extremely poor. The reprocessing model predicts much higher levels of correlation and much shorter lags than those seen in the data, However, we find that (now \textit{unlike} NGC 5548) there does exist a linear transfer function which allows the UV to be produced by the variable X-ray flux (Section~\ref{reverseengineering}). The impulse response function required is broadly distributed between lags of $1.5-20$ days. If this is due to a light travel time, the responding material is co-spatial with the inner broad line region (BLR). We rule out there being optically thick material in a disc geometry on size scales less than $1.5$~light days, as this would produce too much UV fast variability. This variability cannot be hidden by self shielding as proposed for NGC 5548, as even the inner edge of the warm Comptonisation region contributes substantially to the UV flux, so its reprocessing signal would be seen directly, but is not.

We conclude that there is very little evidence for a standard disc structure in NGC 4151 at the time of the intensive monitoring campaign, from either its UV spectrum or lightcurve. Instead, the X-ray spectra are consistent with the inner disc being replaced by a hot flow, and the UV spectra and variability are consistent with being produced by reprocessing in dense gas in the broad line region as first suggested by \cite{KG01}. This is now strongly supported by more recent work (\citealt{CNK18}; \citealt{LGK18}). It remains to be seen whether similar campaigns on objects at higher $L/L_{Edd}$ will give results which connect better to expectations of standard disc models or whether continuum reverberation from inner BLR scales always dominates the UV variability. 

\section{Energetics of Disc Reprocessing \& Soft Excess Comptonisation}
\label{sec:modeling}

The spectra of AGN are complex, typically showing three components which can be modelled by an outer disc emitting the optical and UV continuum, a hot corona producing the hard X-ray power law, with additional emission over and above this below 1~keV (the soft X-ray excess). There is now a growing consensus that much of this soft excess is produced by Comptonisation in warm ($kT_e\sim 0.2$~keV), optically thick ($\tau\sim 10-20$) material which is separate from the hot, optically thin X-ray corona (e.g. \citealt{M98}; \citealt{POP13}; \citealt{M14}; \citealt{MKK15}; \citealt{BRP16}; \citealt{PRM18}). The best current model proposes that this is produced by the accretion energy being released in the upper layers of the disc rather than preferentially in the mid-plane (although there are difficulties with this picture; see \citealt{GKW19}). This produces a sandwich geometry with a heated layer on top of a thin disc of cool, `passive', material (\citealt{POP18}). Gravitational energy is radiated in the upper layers of the disc. Half of these photons are intercepted and thermalised by the underlying passive disc, to then be re-radiated into the warm Compton layer as seed photons. This ties the seed photon luminosity to the warm Comptonisation power, so the shape of the spectrum is fixed by energy balance (\citealt{HM93}; \citealt{POP18}).

Kubota \& Done (2018; hereafter KD18) took this model for the warm Comptonisation and extended it into a fuller picture of the accretion flow: {\tt{agnsed}}. This assumes that the energy released at each radius is set by the Novikov-Thorne disc emissivity (\citealt{NT73}), but that this is only thermalised to a blackbody in the outer disc, where $r>r_{warm}$ (all radii hereafter being dimensionless, where $r=R/R_g$). Further in, at $r_{warm}>r>r_{hot}$, the energy is instead emitted from the warm Compton medium above the passive disc. This warm Comptonisation shape is characterised by its electron temperature, $kT_{warm}$, the spectral index, $\Gamma_{warm}$, and the seed photon temperatures across this zone, which are set by the local gravitational dissipation and hard X-ray illumination. This is unlike the original model of \citealt{POP18}, where the seed photon temperature was a free parameter. Energy balance in a slab geometry hard-wires $\Gamma_{warm}=2.7$, so the only free parameter in the warm Compton shape is $kT_{warm}$. Then there is a final transition to the hot coronal zone for $r_{in}>r>r_{hot}$, which has a vertical scale height $h_{cor}$, and which likely has no underlying optically thick disc. This component's shape is characterised by its electron temperature $kT_{hot}$ and spectral index $\Gamma_{hot}$. The overall normalisation is set by the black hole mass (fixed by either the scaling relations or direct BLR reverberation mapping), the mass accretion rate as a fraction of the Eddington rate, $\dot{m}=L/L_{Edd}$, and the black hole spin.

{\tt{agnsed}} also includes irradiation of the outer blackbody disc and disc-like warm Compton region by the X-ray hot inner region, so that the total flux at each radius $r>r_{hot}$ is $F_{rep}(r) + F_{grav}(r)$ where 
\begin{equation}
F_{rep}(r) = \frac{\text{cos}(n)L_{cor}}{4 \pi (l R_g)^2},
\end{equation}
where $l^2=h_{cor}^2+r^2$ and $\text{cos}(n)=h_{cor}/l$ (\citealt{ZDS99}). The hard coronal flux typically peaks at $100~keV$, so Compton recoil is significant, meaning that reflection cannot be efficient. The maximum albedo is of order $\sim 0.5$ so at least half of the illuminating flux will heat the upper few Thompson optical depths of the 
outer disc and warm Compton region. In the outer disc, this gives rise to blackbody emission with temperature
\begin{equation}
T_{eff}(r) = f_{col}T_{grav}(r)\left(\frac{F_{rep}(r) + F_{grav}(r)}{F_{grav}(r)}\right)^{1/4},
\end{equation}
where the colour temperature correction factor $f_{col}$ is fixed to unity since the temperatures here are low enough that no colour-correction is required (\citealt{D12}). In the warm Compton region, the additional flux input from irradiation adds to the illumination of the underlying passive disc, so modifies the seed photon luminosity and temperature such that $T_{seed}(r) = T_{eff}(r)$ and the spectral index remains fixed at $\Gamma_{warm}=2.7$.

\subsection{Spectral Model Fitting}
\label{sec:modeling_spectral}

\begin{figure}
	\includegraphics[width=\columnwidth]{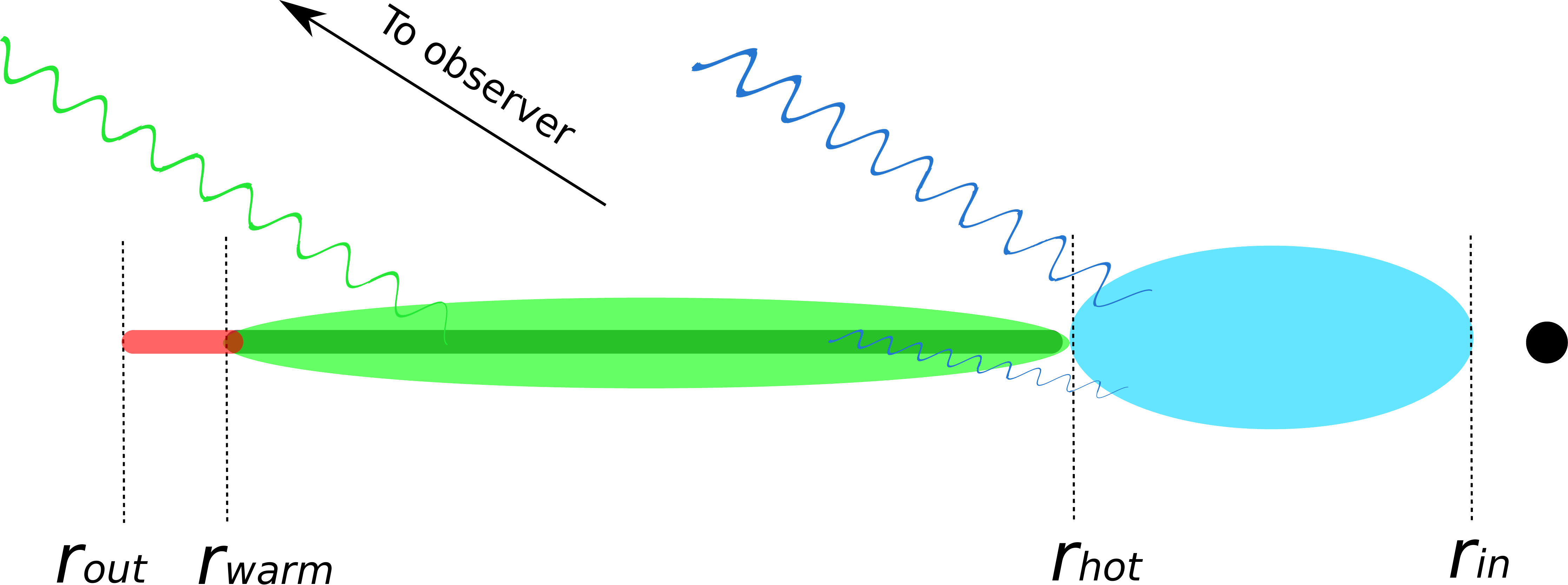}
	\caption{Schematic of the geometry assumed here using {\tt{agnsed}}. The blue region within $r_{hot}$ denotes the hard X-ray corona. The region between $r_{warm}$ and $r_{hot}$ consists of warm, optically thick Comptonising material (green) sandwiching the thin, passive disc (grey). Hard X-rays irradiate the underlying thin disc, in turn changing the seed photon temperature and normalisation of the soft Compton emission from the warm material. Between $r_{warm}$ and $r_{out}$, only cool thermal disc emission is produced (red), found to be negligible for the UVW1 band in Fig.~\ref{fig:rawSED}.}
	\label{fig:STRUCTURE}
\end{figure}

We use {\tt{agnsed}} in {\tt{xspec}} ver. 12.10.0 (\citealt{ABH96}) to model the accretion flow, with this accretion structure sketched in Fig.~\ref{fig:STRUCTURE}. The standard disc (red) extends only from $r_{out}$ to $r_{warm}$. From $r_{warm}$ to $r_{hot}$ the energy is instead released in the upper layers (green), leaving the mid-plane disc as a passive reprocessor (grey). This truncates at $r_{hot}$, leaving the energy to be dissipated in the hot flow (blue).

\begin{figure}
	\includegraphics[width=\columnwidth]{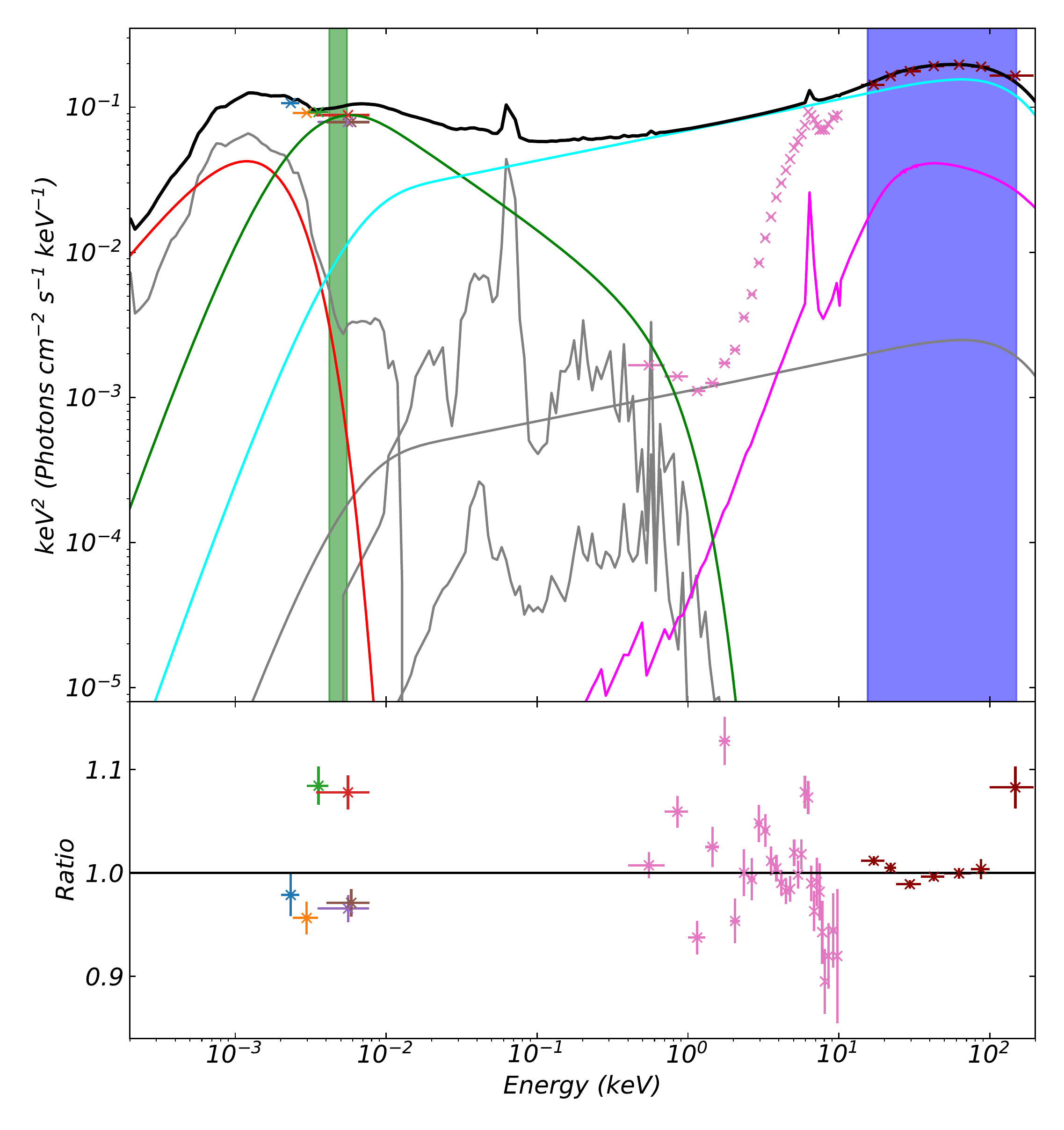}
	\caption{Fit to data with the model {\tt{hostpol + phabs * zredden * [pcfabs * pcfabs * (rdblur * pexmon + agnsed$_1$) + agnsed$_2$ + mekal + mekal]}}. The soft and hard Compton components are denoted by the solid green and cyan lines respectively, while the outer disc component is shown by the solid red line. The pink solid line denotes the distant reflection component. Constant components (stars at optical/UV from an Sb template, and hot gas in soft X-rays which both emits lines and scatters continuum into our line of sight) are shown in gray. The absorbed data are shown as crosses with colours denoting the bandpasses of Edelson et al. (2017; $v$, $b$, $u$, $UVW1$, $UVM2$, $UVW2$, $X1-4$, $BAT$). The green and blue strips denote the UVW1 and \textit{Swift} BAT bandpasses respectively.}
	\label{fig:rawSED}
\end{figure}

The black hole mass is fixed to $M_{BH}=4\times10^7 M_{\odot}$ (as used by E17). We also fix the distance of NGC 4151 to $D=19$~Mpc and assume an inclination angle of $i=53^o$ (i.e. $\text{cos}(i)=0.6$). The electron temperature of the hot Compton component, $kT_{hot}$, is fixed at $100$~keV while its spectral index, $\Gamma_{hot}$, is free. The warm Comptonisation spectral index, $\Gamma_{warm}$, is fixed to 2.7 by the reprocessing in the assumed slab geometry (\citealt{POP18}), while we also fix the temperature, $kT_{warm}$, to $0.2$~keV, as observed (\citealt{GD04b}). 

For the component size scales, we fix the outer disc size at $r_{out}=10^5$, as there is likely to be optically thick material which extends out to torus scales. This assumption matches fairly well with the observed slope of the optical spectrum in the longest wavelength archival data from the Hubble Space Telescope (HST; see Section~\ref{sec:Archival}), but is much larger than the self gravity radius at $r_{sg}=390$. Nonetheless, we expect that material still connects from torus scales to the disc even if it is clumpy rather than smooth. The \textit{relative} normalisation of the thermal disc and two Compton components are then allowed to vary by allowing $r_{hot}$ and $r_{warm}$ to be free parameters, while the overall normalisation of {\tt{agnsed}} is then specified by the (free) mass accretion rate.

However, the SED of NGC 4151 also shows complex absorption and reflection from material surrounding the AGN. \cite{BMD17} perform a careful analysis of the broadband X-ray spectrum and characterize the absorption by two neutral components, both of which are variable: one with $N_H\sim 13-25\times 10^{22}$~cm$^{-2}$ which partially covers $\sim 50$\% of the source, while the other fully covers with $N_H\sim 6-10\times 10^{22}$~cm$^{-2}$. A small fraction of the nuclear flux is scattered around these low ionisation absorbers by gas on large scales, which also emits a mix of photo-ionised and collisionally ionised lines. There is also a clear reflection signature from distant material (a narrow iron K$\alpha$ line core; \citealt{SYW10}), although the existence of inner disc reflection is debated (e.g. \citealt{KBB15}).

Our much more limited \textit{Swift} XRT data do not allow us to constrain all of these components. We instead take our {\tt{agnsed}} continuum and add a single cold reflection component ({\tt{pexmon}}; \citealt{N07}) with spectral index tied to that of $\Gamma_{hot}$, and blurred by low velocities ({\tt{rdblur}}; \citealt{F89}). We also apply two absorption components ({\tt{pcfabs}} + {\tt{pcfabs}}) with column densities $N_H=9\times 10^{22}$~cm$^{-2}$ and $N_H= 25\times 10^{22}$~cm$^{-2}$, and covering fractions $f_{cov}=1$ and $f_{cov}=0.86$ respectively, in order to match the complex curvature. We include a small fraction of the hot coronal emission from {\tt{agnsed}} to model the scattered emission at low energies, together with hot plasma emission to fit the stellar/hot gas lines ({\tt{mekal}}+{\tt{mekal}}; \citealt{MGvdO85}; \citealt{AR92}). All of these components are absorbed by the Galactic column ({\tt{phabs}}) fixed at $N_H\sim 0.02\times 10^{22}$~cm$^{-2}$.

Our data also extend down into the UV and optical so we also include dust extinction ({\tt zredden}), with $E(B-V)$ fixed at $0.03$ from our Galaxy. We also allow for a contribution from host galaxy starlight in the UVOT apertures with an Sb template ({\tt{hostpol}}; \citealt{EGD17}). It is clear that there is substantial host galaxy contamination in the optical photometric bands, but there is much less in the UVW1 (green band in Fig.~\ref{fig:rawSED}). 

The final model is {\tt{hostpol + phabs * zredden * [pcfabs * pcfabs * (rdblur * pexmon + agnsed$_1$) + agnsed$_2$ + mekal + mekal]}}. The {\tt{agnsed$_2$}} term is the scattered continuum, and so contains only the hard Compton component. We fix the normalisation of this scattered emission to $3\%$ of that of the main continuum, for consistency with Beuchert et al. (2017; see their Table 3).

The resulting fit is shown in Fig.~\ref{fig:rawSED}, with the outer disc (red solid line) from $r=10^5-390$, warm Comptonisation (green solid line) from $r=390-90$ and hot Comptonisation (cyan solid line) from $r=90-6$ shown separately. This gives a mass accretion rate of $\text{log}(\dot{m}) = -1.86$, i.e. $1.4\%$ of the Eddington rate. We hereafter refer to this model and parameter set as our `canonical' model.

\begin{figure}
	\includegraphics[width=\columnwidth]{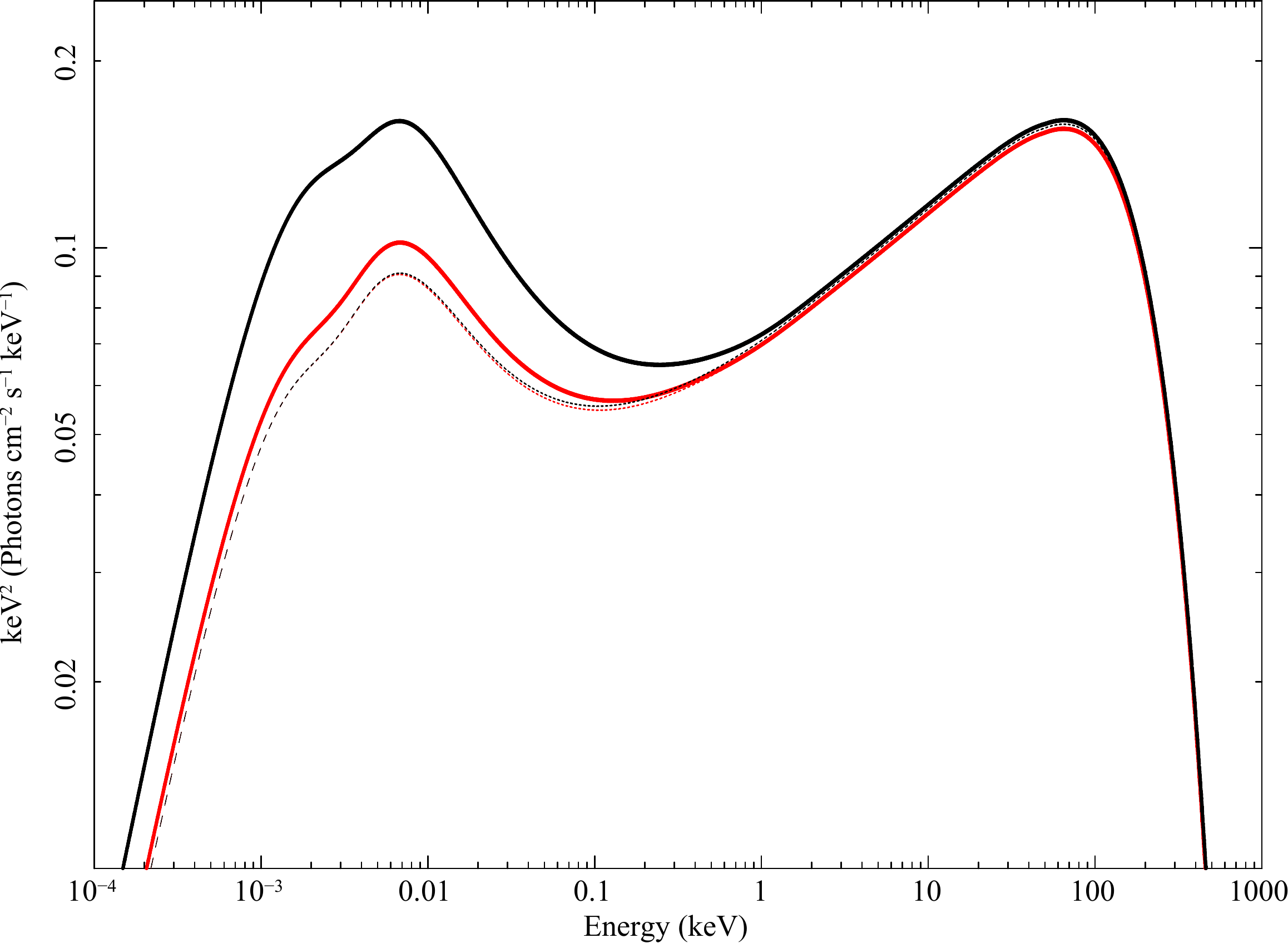}
	\caption{Comparison of different {\tt{agnsed}} parameter sets with reprocessing on (solid lines) or off (dashed lines). The red colour denotes the canonical model we use in this paper, based on the 2016 \textit{Swift} campaign on NGC 4151, with $\text{log}(\dot{m})=-1.86$,~$r_{out}=10^5$, $r_{hot}=90$ and $h_{cor}=10$. The black colour denotes the canonical model as in the red, only now with the scale height of the corona, $h_{cor}$, set to the coronal outer radius, $r_{hot}=90$, instead of the fiducial $h_{cor}=10$.}
	\label{fig:reprocess_on_off}
\end{figure}

The hard Compton corona within $r_{hot}$ is fixed in this modeling to be at a standard scale height of $h_{cor}=10$, as expected from averaging over an optically thin diffuse source with volume emissivity given by gravitational energy release (GD17). We show the effect of this assumption in Fig.~\ref{fig:reprocess_on_off}, where we compare the canonical fit (red, solid line) with one where all parameters are the same except using $h_{cor}=r_{hot}=90$ (black, solid line). Clearly, increasing the scale height of the corona increases the fraction of reprocessing in the optical/UV due to the larger solid angle now subtended by the disc/warm Compton region. More reprocessing will result in a stronger reverberation signal, yet there is a fairly low correlation by eye between the UVW1 and BAT curves in Fig.~\ref{fig:data_curves}. We therefore keep $h_{cor} = 10$ for our canonical model, to minimise the variable component in UVW1 from the hard X-ray illumination. 

For the same reason, we also fix the black hole spin to zero in this fit, thus pushing the ISCO and all characteristic radii as far out as possible. The effect of this is to maximise the disc reprocessing time lag and minimise the disc reprocessing amplitude. The only effect of increasing the spin would be to pull the ISCO and all transition radii inward, resulting in shorter time delays and stronger reprocessing on small scales, which is clearly disfavoured by the shapes of the UV/BAT light curves.

\begin{figure*}
	\includegraphics[width=\textwidth]{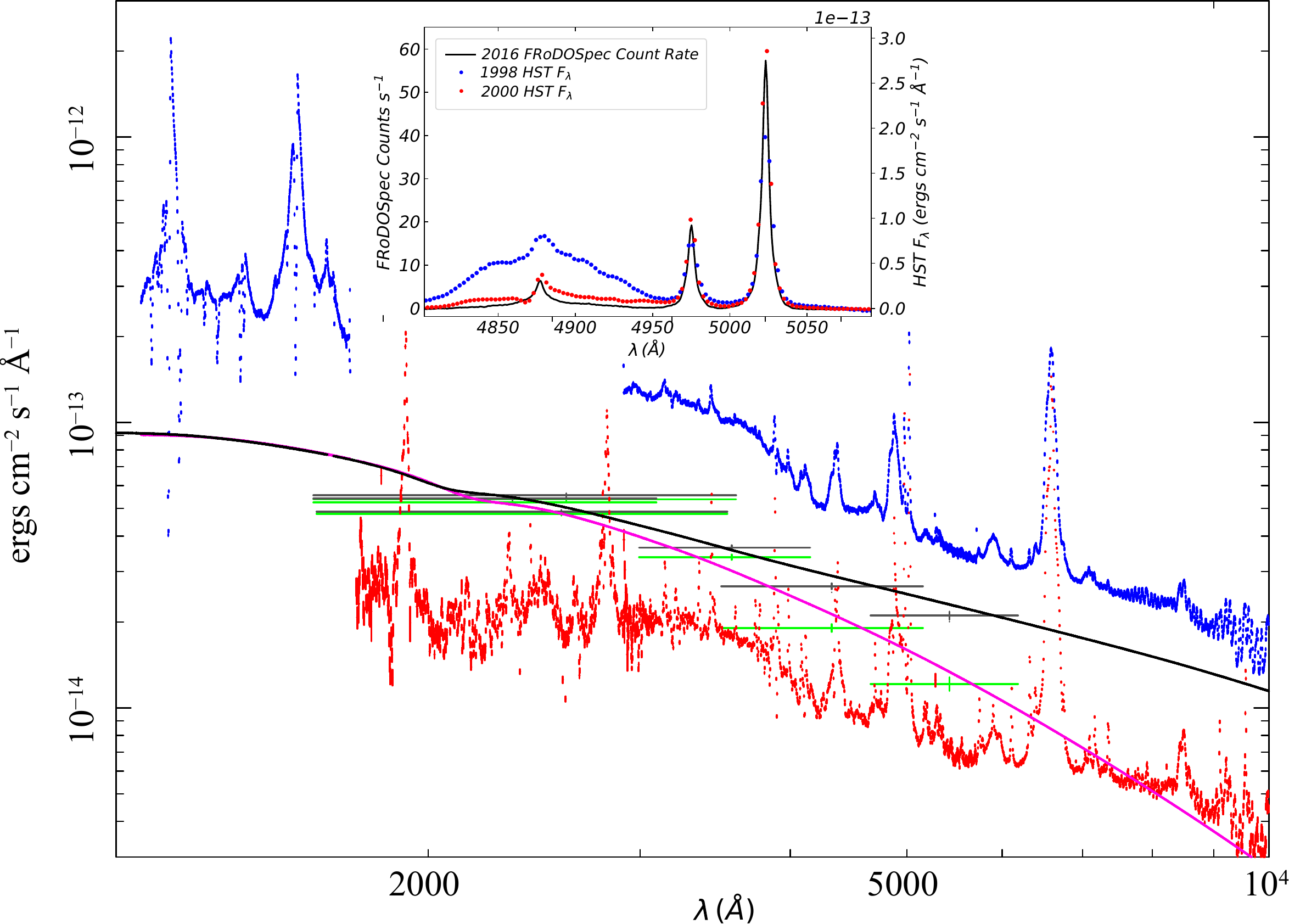}
	\caption{Comparison of HST STIS data taken in February 1998 and May 2000, with the 2016 \textit{Swift} UVOT data which has here been corrected for host galaxy emission. We show HST STIS data taken on 1998-02-10 (blue dots; ObsIDs: O42302070, O42302080, O423020A0) and HST STIS data taken on 2000-05-24 (red dots; ObsIDs: O59701010, O59701020, O59701040). The grey and green data points and errors denote the \textit{Swift} UVOT data with host galaxy emission subtracted according to the predictions of Shapovalova et al. (2008; grey data), and Bentz et al. (2006; green data). The black solid line denotes the canonical model fit to the 2016 UVOT data, with the host galaxy component omitted. The magenta solid line denotes the canonical model fit, but with the outer blackbody disc (as well as the host galaxy) omitted, which is more consisted with the Bentz-corrected data. Inset: Left axis (black line) shows continuum-subtracted Liverpool Telesope FRoDOSpec count rates (ObsID: b_e_20160422_3_2_1_2,dated 22 April 2016), coincident with the \textit{Swift} UVOT campaign. Right axis shows the $H\beta$-OIII line profiles from the HST STIS spectra (1998 again in blue, 2000 in red), also with constant continuum subtractions for comparison of the line profiles. We see a dramatic drop in the broad H$\beta$ line profile between 1998 and both 2000 and 2016, at mildly increased OIII luminosity, indicating a major change in the structure of the BLR between these epochs.}
	\label{fig:1998_2000_2016}
\end{figure*}

\subsection{Comparison to Archival Data}
\label{sec:Archival}

Archival HST spectra give a much better view of the optical/UV continuum. In Fig.~\ref{fig:1998_2000_2016} we show Space Telescope Imaging Spectrograph (STIS) spectra as a function of wavelength collected in both February 1998 (blue dots; ObsIDs: O42302070, O42302080, O423020A0) and May 2000 (red dots; ObsIDs: O59701010, O59701020, O59701040). All of these are taken with 0.1" slit width, so galaxy contamination is negligible, as confirmed by the variability across the entire wavelength range. 

\cite{SPC08} estimate the host galaxy contribution using ground based data at different apertures. They estimate a galactic contribution at $5100~\AA$ of $1.1\times10^{-14}~ergs~cm^{-2}~s^{-1}~\AA^{-1}$ for their $4.2"\times19.8"$ aperture, closest in area to the $5"$ radius circle used by E17. We scale our Sb galaxy template to this value at $5100~\AA$ and subtract the resulting count rates from each UVOT data point resulting in the grey points and error-bars in Fig.~\ref{fig:1998_2000_2016}.

\cite{BDG13} instead used the HST images to estimate the host galaxy contribution in their $5"\times12"$ aperture as $1.7\times10^{-14}~ergs~cm^{-2}~s^{-1}~\AA^{-1}$ at $5100~\AA$. Again scaling our Sb galaxy template to this value at $5100~\AA$ and subtracting the resulting count rates from each UVOT data point yields the green UVOT spectrum in Fig.~\ref{fig:1998_2000_2016}. 

We overlay our canonical fit (black line), which is a fairly good match to the observed slope of the STIS data in both high and low states at wavelengths longer than 5000~$\AA$. This also matches the (grey) UVOT data for the 2016 campaign when the host galaxy continuum derived from \cite{SPC08} has been subtracted. In contrast, the magenta line shows the effect of changing the outer disc radius to the self gravity radius. This is the same as $r_{warm}$, so there is no standard outer disc at all in that model, and we see that it passes through the (green) UVOT data where host subtraction has instead been applied according to the method of \cite{BDG13}. Taking the extrema of possible host galaxy corrections for our data to be given by the Shapovalova- and Bentz-subtracted UVOT data sets (grey and green respectively), we find that models either with or without an outer disc are permitted by our data, although we note that the slope of the longer wavelength STIS data are better matched by models which include an outer disc. Crucially however, both models (and host-subtracted UVOT data sets) converge in the UVW1 band, and so the timing model tested in Section~\ref{sec:modeling2} will be completely unaffected by the level of host galaxy contamination and thus the presence or absence of an outer disc.

\cite{SPC08} also show the long term $5117~\AA$ lightcurve of NGC 4151. They identify low states in 2000-2002 and 2004-2006 (where their data ends) where the AGN flux drops below $3\times 10^{-14}~ergs~cm^{-2}~s^{-1}~\AA^{-1}$. Our dataset is in this category, with a flux of $1-2\times 10^{-14}~ergs~cm^{-2}~s^{-1}~\AA^{-1}$ depending on which host galaxy subtraction is used. Figure 1 in that paper shows a comparison of spectra from 1996 and 2005 which are rather similar to our two HST spectra in terms of their continua. These show that the broad base of the H$\beta$ line has reduced by much more than the continuum flux change in these low states. While simultaneous HST spectra are not available for the 2016 \textit{Swift} campaign period, fortunately there are Liverpool Telescope FRoDOSpec data taken coincident with the 2016 \textit{Swift} monitoring. While these cannot easily be flux calibrated, a continuum-subtracted spectrum around H$\beta$ is shown in the inset on Fig.~\ref{fig:1998_2000_2016} in black. The broad component is low, similarly to those seen in the low states of \citealt{SPC08}. Comparison with the 1998 HST $F_{\lambda}$ spectrum (also shown in the inset and continuum subtracted) confirms that the 2016 $H\beta$ line profile was much lower than in 1998, while the OIII flux was mildly higher. The size of this change in $H\beta$ could not be accounted for by a simple flux calibration, indicating that there was likely a major change in the structure of the broad-line region between the 1998 and 2016 epochs; this suggests that NGC 4151 may have been at the edge of - or even undergoing - a `changing state' episode at the time of the 2016 campaign. The same can also likely be said for the 2000 epoch, which shows a very similar line profile to that in 2016. This is all most likely real variability rather than simply changes in line of sight obscuration, as \cite{LZW10} show that the optical continuum correlates with the hard X-rays over these long timescales, while \cite{KHA13} also confirm that the infrared follows the optical continuum change.

\section{Light Curves from Warm Comptonisation of Disc-Reprocessed Photons}

\label{sec:modeling2}
\begin{figure}
	\includegraphics[width=\columnwidth]{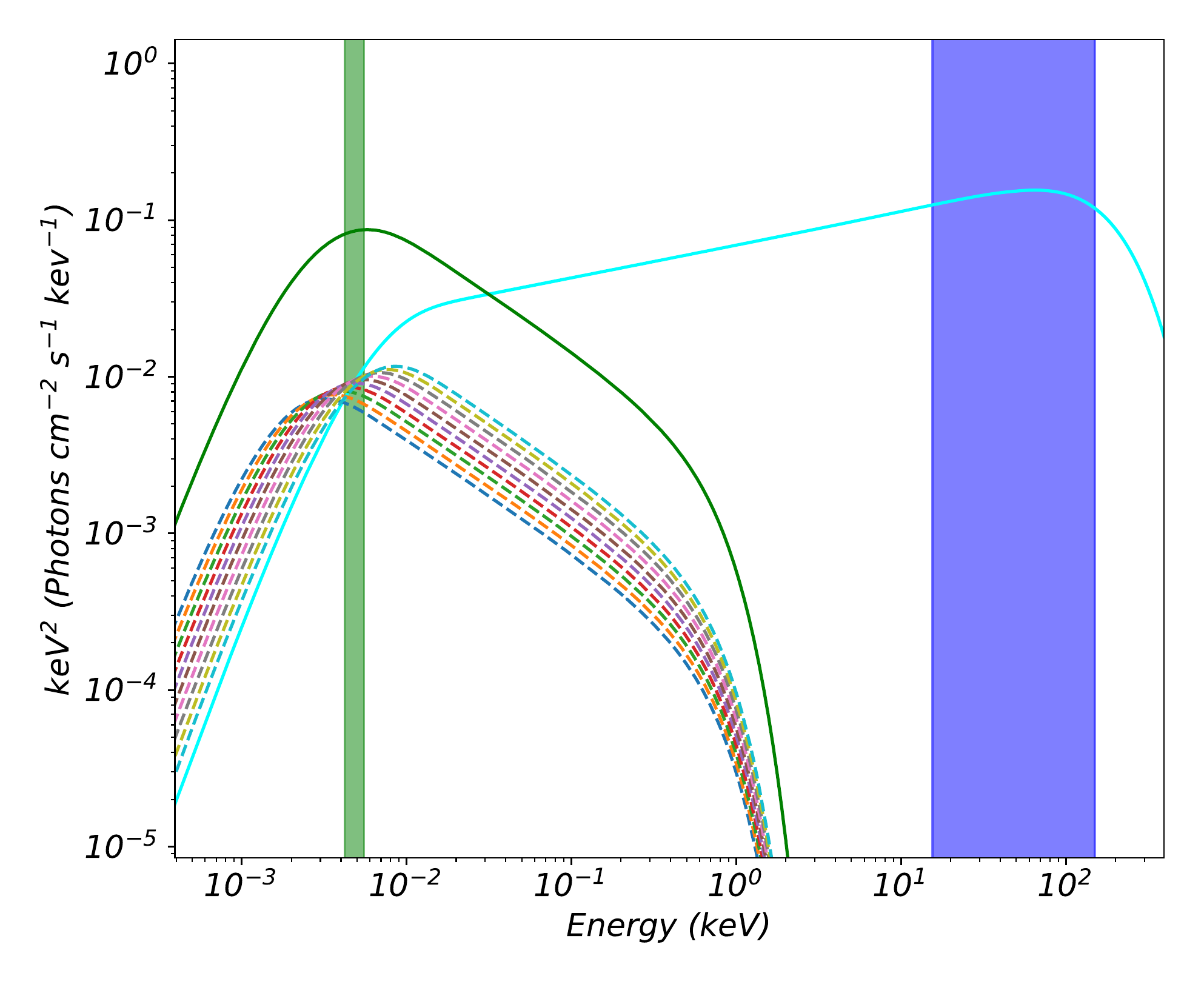}
	\caption{Model spectrum: Cyan line denotes the total hard Compton emission from the corona, which illuminates the disc. The solid green line shows the total soft-Compton emission from the warm material on the disc, which reprocesses all thermal emission from the underlying thin disc. The dashed lines show the soft-Compton emission from the 10 constituent annuli in the soft-Compton region. The seed photon temperature (the blackbody temperature of the underlying disc) and normalisation of these annuli are modulated by the illuminating hard X-ray emission. The green and blue bands denote the UVW1 and \textit{Swift} BAT bandpasses respectively.}
	\label{fig:SED}
\end{figure}

We now take the soft Compton and hard X-ray components from the canonical model of the previous section, and follow a similar procedure to GD17 in order to extract predicted lightcurves in the UV. We choose here to use only the UVW1 band as there is very little host galaxy contamination in this band. Since there is also little contribution from the outer standard disc in this bandpass (see Fig.~\ref{fig:1998_2000_2016}), we model reverberation only from the warm Comptonisation zone, for clarity. Tests performed \textit{a posteriori} indeed confirm that inclusion of the outer disc reverberation has no impact on the predicted light curves, because the temperature at which the emission thermalises at large radii is too low to contribute to UVW1. In Fig.~\ref{fig:SED}, the soft Compton spectra for 10 annuli spanning the full range of warm zone radii are shown as dashed lines; it is the sum of these spectra over \textit{all} the passive disc/warm Compton annuli which produces the overall warm Comptonisation spectrum. We now quantify how these warm Compton annuli respond to variability in the hard X-ray flux.

The Impulse Response Function ($IRF$) accounts for the light travel times to different radii and azimuthal angles within a given annulus, and so for a given radius, this function describes what fraction of the reprocessed flux from that radius has time delay $\tau$ with respect to the illuminating continuum. We calculate this as in GD17 (based on \citealt{WH91}), giving the reprocessed flux from an annulus at radius $r$ and time $t$ as
\begin{equation}
F_{rep}(r, t) = \frac{\text{cos}(n)}{4 \pi (l R_g)^2} \int^{\tau_{max}}_{\tau_{min}} IRF(r,\,\tau)\,L_{\,cor}(t-\tau)\,d\tau.
\end{equation}
This causes the effective temperature of the seed photons supplying the warm Comptonisation in that annulus to vary as
\begin{equation}
T_{seed}(r,\,t) = f_{col}T_{grav}(r)\left(\frac{F_{rep}(r,\,t) + F_{grav}(r)}{F_{grav}(r)}\right)^{1/4}.
\end{equation}
We stress again that these equations assume a `passive' underlying disc, i.e. that there is no intrinsic variability in the underlying disc. The only source of variability in the warm Comptonisation here is from external heating by the illuminating hard X-rays. We assume that this change in heating is followed quickly by the change in cooling via reprocessing in the passive mid-plane material, so that the spectral index and temperature of the spectrum from a given annulus does not change, while its seed photon temperature does change. By computing the seed photon temperature at each time-step for each annulus, our code not only accounts for the changing normalisation of the overall warm Comptonisation due to the driving continuum, but also for the changing \textit{shape} of this  component as different annuli are illuminated, changing their seed photon temperatures. We therefore stress that no single radius dictates the optical/UV luminosity.

The spectral model assumes that the seed photons for the hard X-rays are from the warm Comptonisation zone, so the hot corona emission extends down into (and hence directly contributes to) the UVW1 band (see Fig.~\ref{fig:SED}). We include this direct variable emission, so the UVW1 lightcurve is the sum of the contributions in this band from the primary hard X-ray source, the constant, intrinsic dissipation in the warm Comptonisation region, and the variable reprocessed flux. 

\begin{figure}
	\includegraphics[width=\columnwidth]{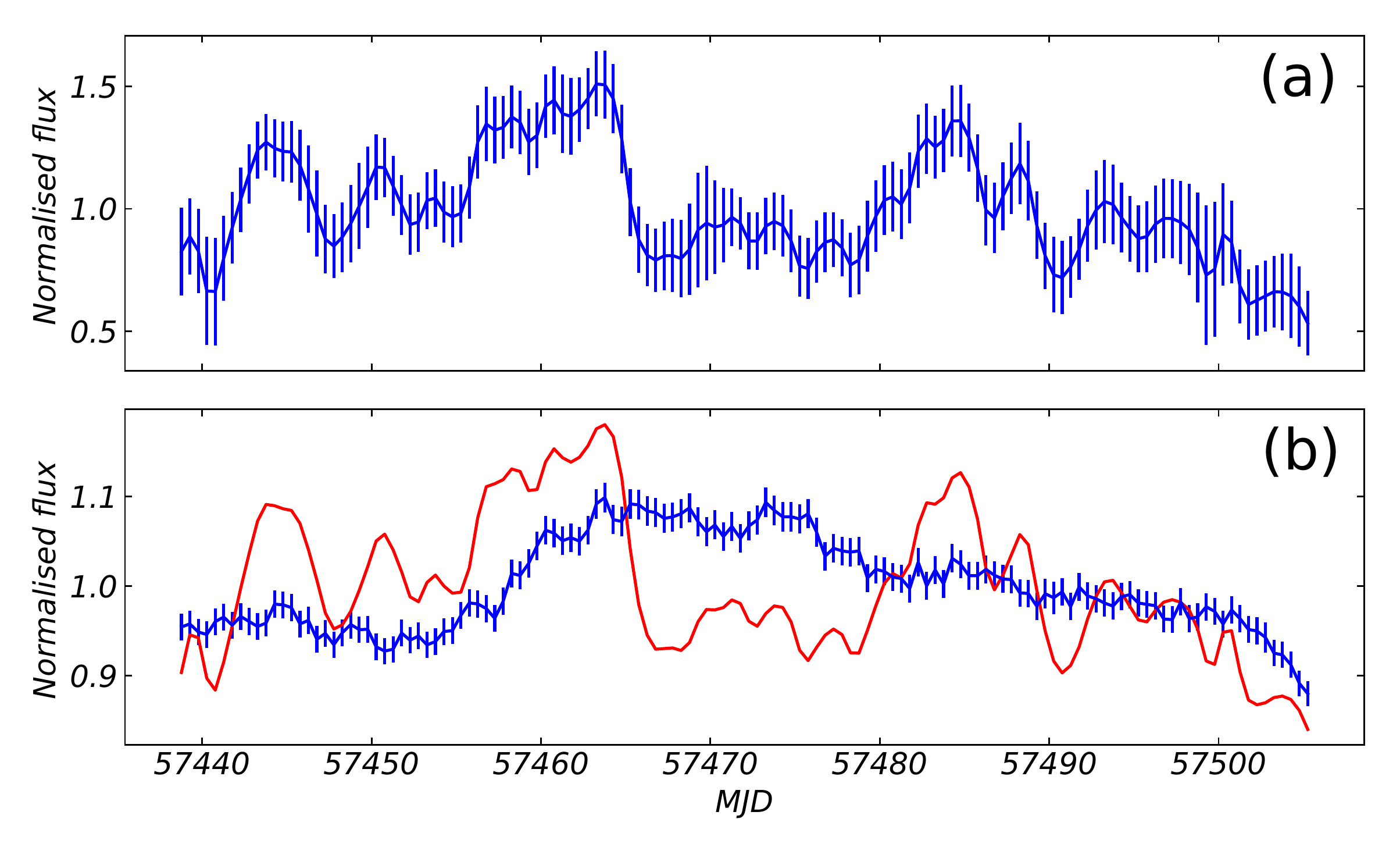}
	\caption{Top panel: Re-binned and interpolated \textit{Swift} BAT light curve. Bottom panel: Re-binned and interpolated UVW1 light curve in blue. The predicted UVW1 light curve resulting from the model assuming disc illumination driving warm Comptonisation is denoted in red. The lack of agreement between the data and model are clear.}
	\label{fig:lightcurves}
\end{figure}

Rather than simulating the driving hard X-ray fluctuations, we use the observed \textit{Swift} BAT light curve of NGC 4151 (shown in E17 and in Fig.~\ref{fig:data_curves}a) as the driving lightcurve. We interpolate this on a grid of $0.05$~days, and then re-sample it to produce give an even sampling of $dt=0.5$~days, shown in Fig.~\ref{fig:lightcurves}a. This light curve is fed into our disc reprocessing-warm Comptonisation model, allowing us to predict the UVW1 light curve from our physical model of the accretion geometry.

Fig.~\ref{fig:lightcurves}b shows our resulting UVW1 light curve (red), compared to that observed (blue). Our simulated UVW1 light curve is very clearly not like the real UVW1 data, with much more fast variability than is observed. In particular, the peaks at MJDs 57460 and 57485 in the modeled UVW1 curve are completely absent in the observed UVW1 curve, replaced instead by a single broad peak centered at MJD 57470.

\begin{figure}
	\includegraphics[width=\columnwidth]{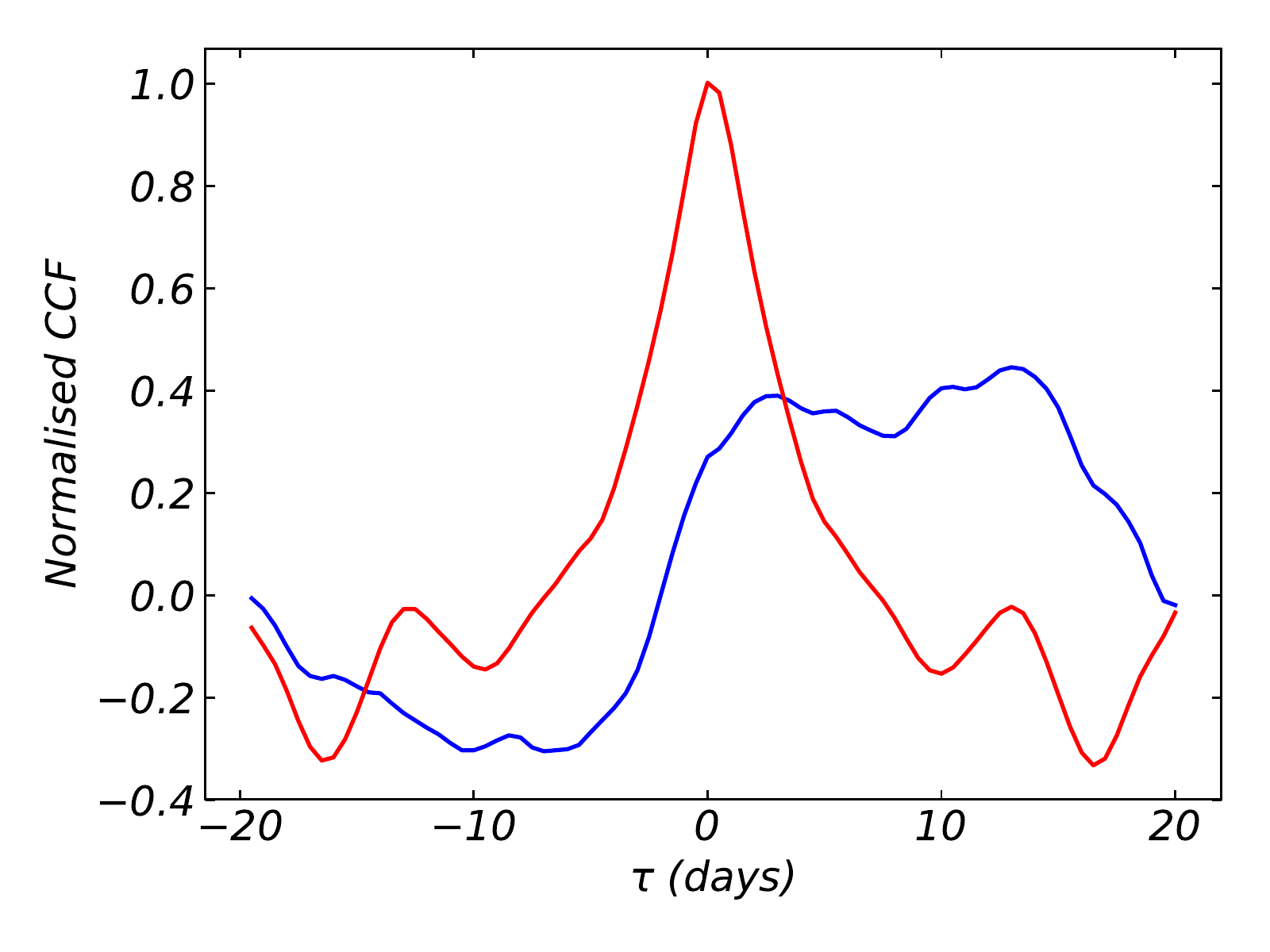}
	\caption{Normalised cross-correlation functions between the \textit{Swift} BAT light curve and UVW1 observed light curve (blue), and the \textit{Swift} BAT light curve and modeled UVW1 light curve (red).}
	\label{fig:CCFs}
\end{figure}

The discrepancy is even clearer in Fig.~\ref{fig:CCFs}, showing the cross-correlation functions between the hard X-ray curve and the modeled (solid red line) or observed (solid blue line) UVW1 curves. These CCFs are computed according to the procedure of GD17, which accounts for the error bar variance. It is clear that the predicted lag of $\sim 1$ day for the simple reprocessing/warm Comptonisation model where the Comptonisation takes place between $90$ and $390$~$R_g$ is far shorter than that observed, and so the model predicts far more correlation between the BAT and UVW1 light curves than observed. Instead, the data shows much less correlation, with positive lags between $2$ and $15$ days. E17 calculate that the \textit{Swift} BAT band and UVW2 bands are correlated with a significance of only $72\%$ when the red-noise character of AGN variability is taken into account. We must therefore conclude that the processes generating the emission in each of these bands are, at best, weakly related, unlike the extremely strong correlation between all the UVOT bands (E17).

We test the robustness of these timing model results to the input spectral model in Appendix~\ref{truncated_disc}. There we show an extreme model, where including additional reddening allows the intrinsic SED of these data to be fit with \textit{no} warm Compton zone. In this case, $r_{warm}=r_{hot}=60-70~R_g$, so the UV emitting material is even closer, resulting in an even stronger and faster reverberation signal which is even more inconsistent with the observed variability. We conclude that if the UV is powered by optically thick emitting material then it must be present at radii significantly below $1$~light-day ($440~R_g$), irrespective of whether it emits as blackbody or warm thermal Comptonisation. Yet such material in a disc-like geometry is strongly ruled out by the variability. The results presented in this section are therefore robust to the SED decomposition, as all reprocessing taking place in this class of models must be on sub-day time-scales.

\subsection{The disconnect between the hard X-rays and the UV continuum}
\label{sec:Implications}

Our spectral model gives a good description of the broad band continuum in NGC 4151 during this campaign, where the UV continuum comes from a region of warm Comptonising material. It also predicts the UVW1 variability from the assumed disc geometry and emissivity, which place this material at $r=90-390$. However this picture is wholly ruled out by the observed UVW1 light curve, which exhibits much less of the fast variability predicted by the model in Fig.~\ref{fig:lightcurves}, and is instead dominated by variability on longer timescales, up to 15 days (Fig. 8).

Instead, we conclude that our model of the accretion flow is wrong; it overestimates the amount of fast variability, and underestimates the slow variability. This all points to there being an additional spectral component in UVW1, which is slowly variable, and which replaces part of the fast-responding outer disc/warm Comptonisation in the current spectral decomposition. Fig.~\ref{fig:CCFs} shows that there is a broad lag signature from $2-15$~days, a timescale which neatly encompasses the size scale of the BLR in this source, with HeII measured at $2-3$~light days, H$\beta$ at $6$~light days and H$\alpha$ at $11$~light days (\citealt{PH04}), all of which are substantially smaller than the IR torus size scale of $40$~light days (\citealt{OMT14}).

A potential contribution from diffuse emission from the BLR was first identified by \cite{KG01}, and given strong impetus by the intensive monitoring campaign results which pointed to longer response time scales than expected from the disc (E17, see their Fig.~5). Similar results are also seen in the sparser data of \cite{CNK18}, where they derive a spectrum of the slowly reverberating component in the optical and identify this with the BLR. \cite{LGK18} show predictions for the diffuse BLR flux for clouds of different densities. It is mainly composed of a series of radiative recombination continua - most obviously at the Balmer and Paschen limit - as well as free-free continua (as was suggested by \citealt{B87}!). This non-blackbody shape is important, as a key problem with blackbody thermalisation is that this gives temperatures which are too low to contribute to UVW1 for the large size scales required by the lags (GD17). Conversely, material at high enough temperature to contribute blackbody radiation to UVW1 would have too small an area for the required luminosity (GD17).

\section{Recovering the Impulse Response Function}
\label{reverseengineering}

\begin{figure*}
	\includegraphics[width=\textwidth]{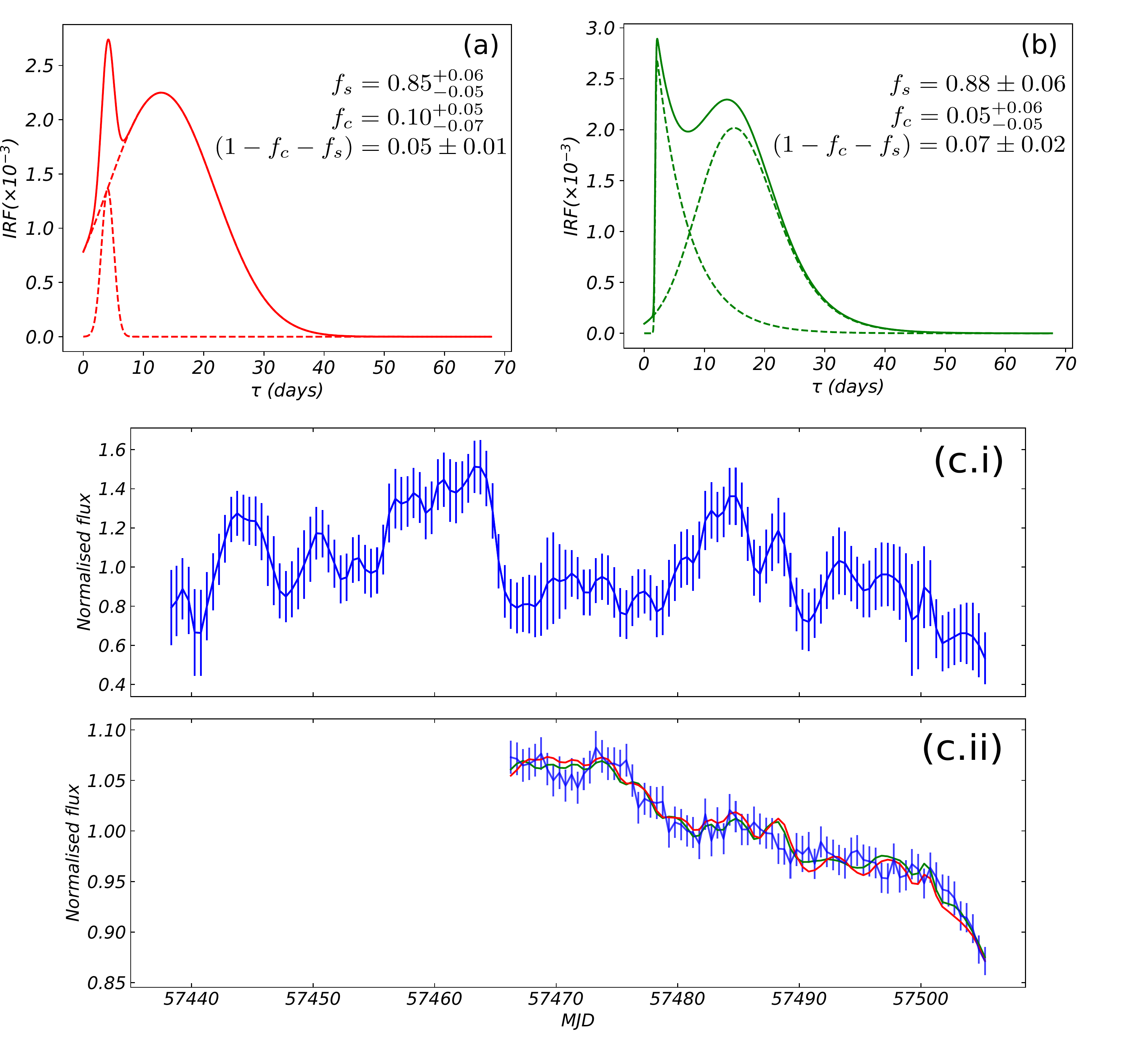}
	\caption{Modeled UVW1 light curve using phenomenological impulse responses to the BAT curve. Panel~(a): Red, solid line denotes the impulse response function used to fit the UVW1 light curve, composed of a sum of two Gaussians. The dashed lines denote the two constituent Gaussians which compose this function. Panel~(b): More sophisticated impulse response function used to fit the UVW1 light curve, with functional form in equation~(\ref{eq:fiducialIRF2}). Panel~(c.i): Re-binned and interpolated \textit{Swift} BAT light curve. Panel~(c.ii): Re-binned and interpolated observed UVW1 light curve in blue. Modeled UVW1 light curves denoted in red and green, where the UV emitting component is directly modulated by the \textit{Swift} BAT lightcurve convolved with either the $IRF$ of panel~(a)~(red) or panel~(b)~(green).}
	\label{fig:reverseengineered}
\end{figure*}

The conclusion that there is an additional optical/UV component from the BLR means that the accretion continuum contribution in our spectral fit is overestimated. Here we attempt to separate out the contribution of this new component. We assume that the UVW1 flux can be split into a constant component, carrying a fraction $f_c$ of the average UVW1 flux, a slowly variable fraction due to reprocessing from large scales, $f_s$, while the remaining fraction, $(1-f_c-f_s)$, is fast-varying from either a direct contribution of the hard X-rays in the UVW1 band its reprocessing on the optically thick disc/warm Compton (lagged and smoothed by the light travel time of less than $1.5$~days). For simplicity we combine both of these fast variability components together. This means that the UVW1 emission is described by
\begin{equation}
\label{FUVW1}
\begin{split}
\frac{F_{UVW1}(t)}{<F_{UVW1}(t)>} = f_c + & f_s IRF(t) \circledast \frac{F_{BAT}(t)}{<F_{BAT}(t)>}  \\
	     &+ (1-f_c-f_s) \frac{F_{BAT}(t)}{<F_{BAT}(t)>}.
\end{split}
\end{equation}

We first assume that the slow response function is a Gaussian, so $IRF(\tau) = e^{\frac{(\tau-\mu)^2}{2\sigma^2}}$ where $\mu$ is the centroid and $\sigma^2$ is the variance. The observed slow response has a contribution up to $\sim20$ days (see Fig.~\ref{fig:CCFs}) so we calculate the $IRF$ out to 30 days. This means that we have to discard the initial 30 days of the UVW1 lightcurve, as it depends on the X-ray lightcurve prior to the monitoring campaign. We calculate the resulting truncated UVW1 lightcurve, minimising $\chi^2$ between our predicted lightcurve and the data to determine the best fit $f_c$, $f_s$ and $IRF$ parameters. However for a single Gaussian $IRF$ this gives an unacceptable fit ($\chi^2/d.o.f.=316/75$), so we instead use the observed double peaked CCF (Fig.~\ref{fig:CCFs}) to guide our choice of a double-Gaussian for the IRF. This requires three additional parameters (a relative normalisation, $K_2$, as well as the additional Gaussian centroid and variance). This now gives a good description of the lightcurve, with the model shown as the red line in Fig.~\ref{fig:reverseengineered}c.ii, with associated $IRF$ in Fig.~\ref{fig:reverseengineered}a, and $\chi^2/d.o.f.=52/72$. 

We also test a more sophisticated $IRF$ with more free parameters. Motivated by the shape of the observed CCF in (Fig.~\ref{fig:CCFs}) we replace the Gaussian functions with exponentially modified Gaussians, characterised by rates $\lambda_1$ and $\lambda_2$, so that
\begin{equation}
\label{eq:fiducialIRF2}
\begin{split}
IRF(\tau)&= I\left[e^{\frac{\lambda_1}{2} \left(2\mu_1 +\lambda_1\sigma_1^2 - 2\tau \right)}\,\text{erfc}\left(\frac{\mu_1+\lambda_1\sigma_1^2-\tau}{\sqrt{2}\sigma_1}\right) \right. \\
&\left.+ K_2 e^{\frac{\lambda_2}{2} \left(2\mu_2 +\lambda_2\sigma_2^2 - 2\tau \right)}\,\text{erfc}\left(\frac{\mu_2+\lambda_1\sigma_2^2-\tau}{\sqrt{2}\sigma_2}\right)\right].
\end{split}
\end{equation}
Here $erfc$ is the complementary error function, and $I$ normalises the $IRF$ such that its integral is unity as required. This gives a better fit to the UVW1 lightcurve (IRF in Fig.~\ref{fig:reverseengineered}b green line in Fig.~\ref{fig:reverseengineered}c.ii), but only by $\Delta\chi^2=10$ for the two additional free parameters. Over-fitting the data is clearly an issue, but we show this in order to show the size of systematic uncertainty on the IRF, $f_c$ and $f_s$.

\begin{figure}
	\includegraphics[width=\columnwidth]{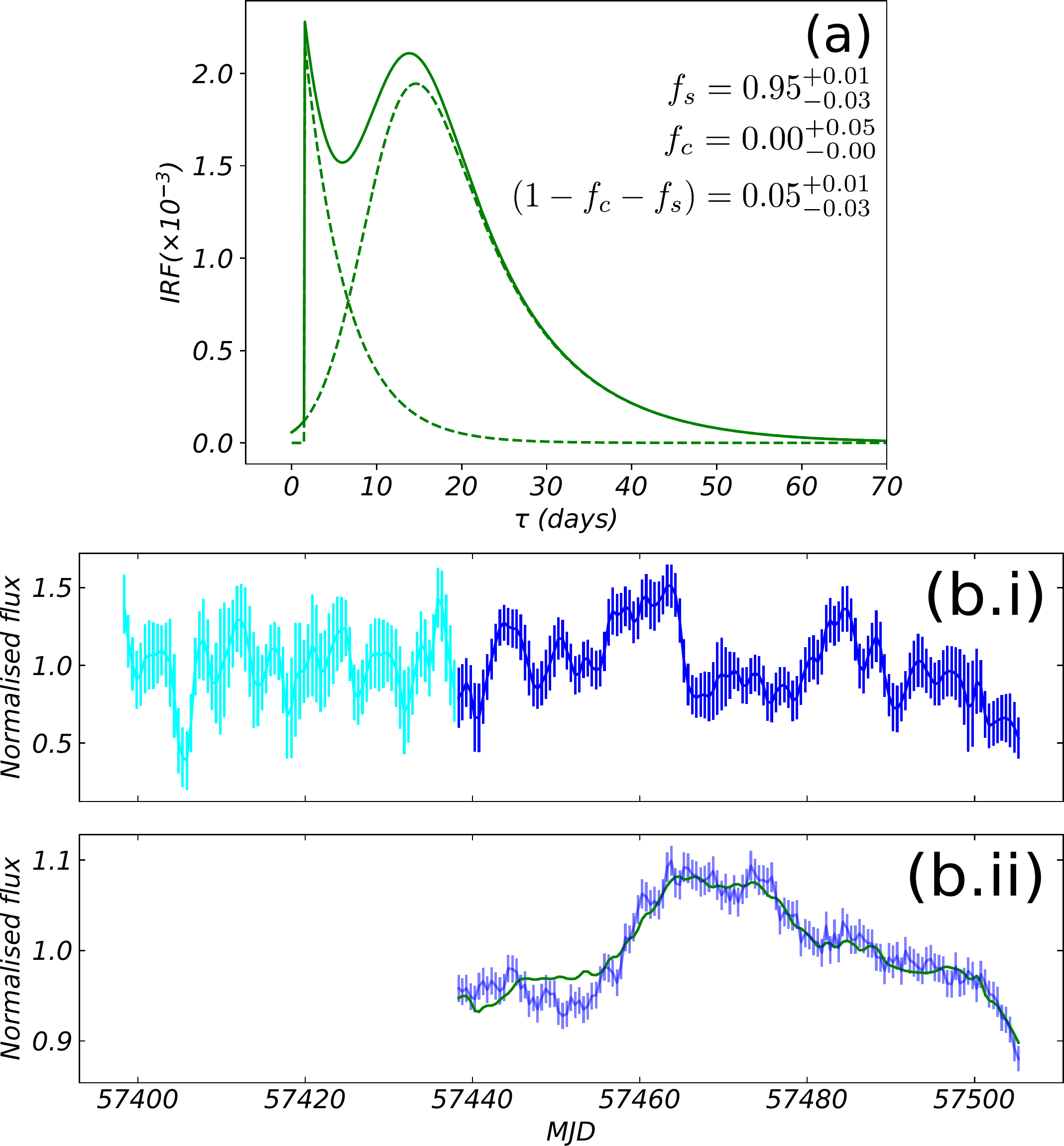}
	\caption{Modeled UVW1 light curve using exponentially-skewed Gaussian response to the BAT curve input, for the entirety of the pointed campaign. Panel~(a): $IRF$ used to fit the UVW1 light curve, with functional form in equation~(\ref{eq:fiducialIRF2}). Panel~(b.i): Re-binned and interpolated \textit{Swift} BAT light curve from the 2016 pointed campaign denoted in blue. BAT light curve prior to the pointed campaign is shown in cyan. The inclusion of this data, covering $40$~days prior to the beginning of the 2016 campaign allows us to make a prediction for the entire UVW1 curve from the beginning of the pointed campaign at MJD 57435 onward. Panel~(b.ii): Re-binned and interpolated observed UVW1 light curve in blue. Modeled UVW1 light curve denoted in green, where the UV emitting component is directly modulated by the extended \textit{Swift} BAT lightcurve of panel~(b.i), convolved with the $IRF$ of panel~(a).}
	\label{fig:irf_fake_earlyXrays}
\end{figure}

To test the validity of the derived fractions and IRF, we now include $40$~days of UVW1 data prior to the beginning of the 2016 campaign by extending the \textit{Swift} BAT lightcurve back in time using the standard all sky monitoring data. Since the \textit{Swift} BAT instrument is a large sky monitor (covering a quarter of the sky at any given time), we can do this without sacrificing signal-to-noise at earlier times. We then interpolate within this lightcurve to produce an X-ray lightcurve on an equivalent time grid to the 2016 campaign data, facilitating the $IRF$ fit. This extension of the BAT data to earlier times allows us to compare our predictions to the most important feature in the UVW1 lightcurve, namely the large rise around MJD 57460. We re-run both $IRF$ models to fit the entire UVW1 lightcurve. The exponentially modified double Gaussian $IRF$ yields $\Delta\chi^2=30$ better than the double Gaussian case, so we show this alone in Fig.~\ref{fig:irf_fake_earlyXrays}. The fit now gives a good description of the main features of the entire UVW1 lightcurve ($\chi^2/d.o.f.=145/126$).

\subsection{Physical Origin of the Derived IRF}
\label{PhysicalOrigin}

The IRFs we have derived in Section~\ref{reverseengineering} show that the slow variability can be matched by reverberation, with $90\%$ of the UVW1 flux predominantly coming from BLR size scales of $1.5-20$ light days. Only $\sim5-10\%$ of the UVW1 band emission can be correlated with the hard X-rays on short timescales. This is a factor $3$ smaller than that predicted by the model explored in Section~\ref{sec:modeling2}, but is consistent with the $9\%$ of UVW1 flux predicted to be contributed directly by the low energy extension of the hard X-ray component. The hard Compton component shape - well constrained by the peak in \textit{Swift} BAT - requires that its seed photons come from UV energies or below, and so this hard component contribution to UVW1 must be present. This leaves very little room for any UVW1 contribution from warm Comptonisation or thermal disc emission from within $1.5$~light days ($\lesssim 650~R_g$). The accretion structure within $1.5$~light days is therefore far more consistent with being a pure radiatively-inefficient accretion flow (RIAF; \citealt{YQN03}), where seed photons are generated internally by cyclo-synchrotron emission (\citealt{V16}; \citealt{ID18}).

\begin{figure*}
	\includegraphics[width=\textwidth]{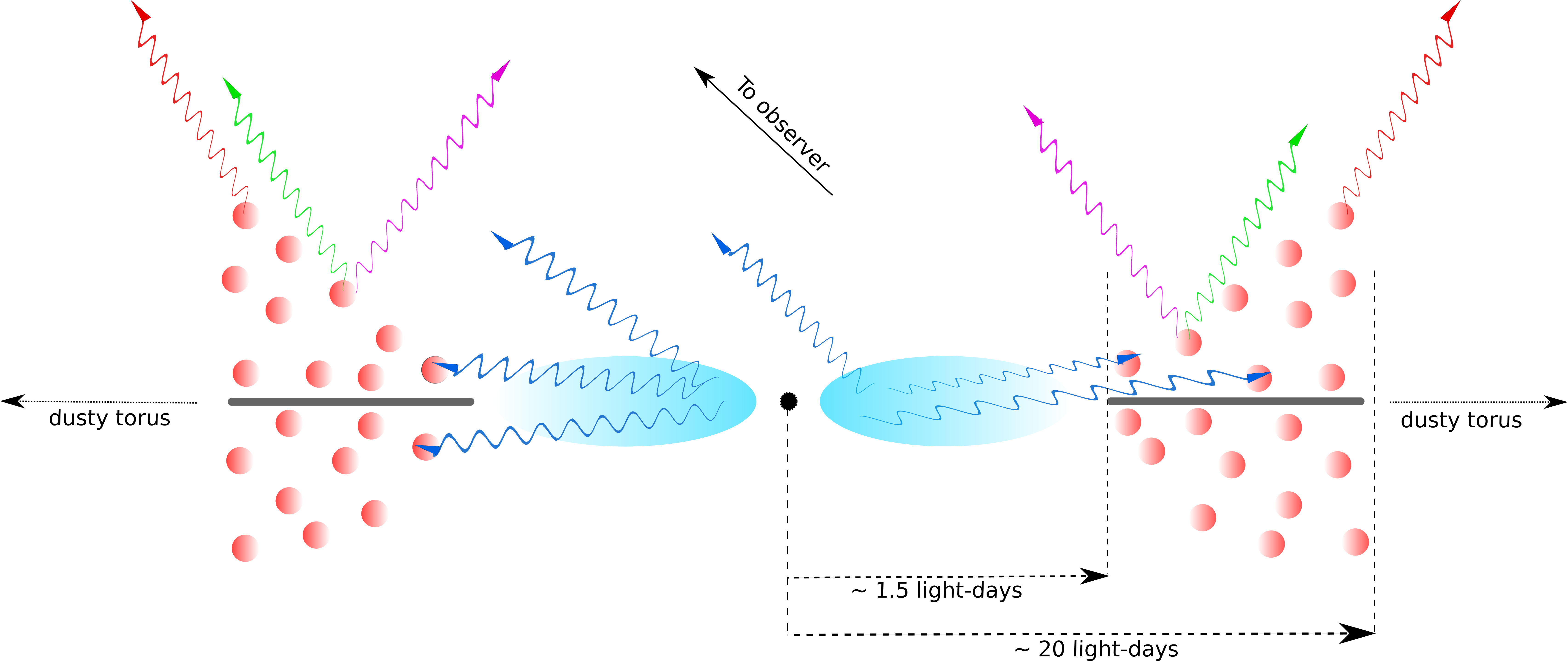}
	\caption{The geometry we propose to explain both the observed SED and derived impulse response functions. The blue zones denote the radiatively inefficient accretion flow which emits strongly in the hard X-rays. The red/white circles denote the broad-line region clumps which reprocess the hard X-ray emission and on which line emission is excited. The grey zones denote outer disc material from which the reprocessor clouds may be launched via e.g. a dust driven wind. The existence of this disc material down to $\sim 500$~$R_g$ has been suggested by \citealt{MCZ17} from the velocity width of the Fe~K$\alpha$ line. Wiggly lines denote photons in the optical (red), UV (green) and hard X-rays (blue) bands, and line emission (purple) which is preferentially produced on the illuminated (white) side of the BLR clumps.}
	\label{fig:STRUCTURE2}
\end{figure*}

We reiterate that in this picture, the lack of short ($<1$~light-day) response is due to there being little-to-no optically thick material on these scales. We illustrate this geometry in Fig.~\ref{fig:STRUCTURE2}, where the inner disc has evaporated into a hot, optically thin flow which extends out to $\sim600R_g$. This result is indeed corroborated by the very recent work of \cite{ZMC19}, who find no evidence for a relativistically broadened iron line component after accounting for absorption effects, and thus no evidence for an inner disc.

\cite{LGK18} performed detailed calculations of diffuse emission from the BLR. They show that while `standard' BLR cloud densities of $n_H\sim10^{10}$~cm$^{-3}$ efficiently produce the observed lines, higher density clouds emit continuum (via free-free and bound-free interactions) more efficiently than lines. They show that a population of denser clouds within the BLR produce a substantial diffuse continuum. Additional evidence for these clouds comes from BLR reverberation studies of NGC 5548. The emission line `holiday', where the high ionisation BLR lines change their response to the continuum variability, can be explained by changes in dense clouds on the inner edge of the BLR filtering the continuum seen by the BLR (\citealt{DFK19}; \citealt{DFP19}). In both NGC 4151 and in NGC 5548, this material is probably the same as that which produces the time variable X-ray absorption. This is seen as a blue-shifted absorption line against the broad H$\beta$ emission, indicating that it is part of a dense outflow; as the H$\beta$ line is not a resonance transition, absorption requires substantial collisional excitation to the $n=2$ level. \cite{DFP19} show again that the ratio of continuum to lines increases as the density increases, with densities higher than $n_H \sim 10^{12}$~cm$^{-3}$ producing substantial optical/UV continuum (both free-free and Balmer/Paschen recombination continua; see their Fig.~3). If the density becomes too high however, the emission thermalises to a temperature of $T<L_{cor}/(4\pi R^2 \sigma_{SB})\sim3500$~K - too low to contribute to the UV band.

In this picture, the double response $IRF$ shape we see in Fig.~\ref{fig:reverseengineered}b does not describe two separate mechanisms, but the distributed response of partially ionised clouds, from the near and far side of the AGN. This is very similar to the geometry invoked for the clumpy molecular torus material, but on smaller scales, and we note the similarity of our $IRF$ to those of the infrared reverberation simulations of Almeyda et al. (2017; see e.g. their Fig. 7).

To check whether BLR reprocessing of hot flow emission is energetically feasible, we can consider the energy absorbed by the obscurers in the canonical SED fit. By computing the hard coronal luminosity before and after obscuration, we find that 35\% of the hot coronal power should be absorbed by the obscurers in the canonical SED decomposition. According to HST extended narrow-line region studies (see e.g. \citealt{KBH00}), the narrow-line region cone in NGC 4151 has an opening angle of approximately $60^o$. The BLR should fill in the remainder of the sky, so that it subtends a solid angle of around $11$~sr on the hard corona, or 86\% of the sky from the perspective of the central engine. This would imply that $0.86*35 = 30\%$ of the hard coronal power is absorbed by the BLR, or the wind at its inner edge. Of course since the density of the obscurer is likely to be higher closer to the disc plane, this intercepted (and reprocessed) power is actually likely to be higher, but it is already sufficient to reproduce the power in the warm Compton SED component, which is $L_{warm}/L_{hard} = 27.5\%$ of the hard Compton luminosity. There can also be additional power from reflection of the higher energy flux, so reprocessing can work energetically.

The fraction of flux which is constant is $f_c < 0.05$, according to the error ranges shown in Fig.~\ref{fig:irf_fake_earlyXrays}. We calculate the contribution from the host galaxy in the SED model (Section~\ref{sec:modeling_spectral}). This is $\sim 3\%$ of the UVW1 band in the canonical SED fit, which is fully consistent with the contribution preferred by the exponentially modified Gaussian $IRF$ fit to the final lightcurve ($f_c = 0^{+0.05}_{-0.00}$). A larger constant UV flux is only allowed in the models which do not include the first $40$~days of data (Fig.~\ref{fig:reverseengineered}), but we note that any constant UV disc contribution invoked to explain these higher levels of $f_c$ must be accompanied by a comparable level of fast reprocessing (as established by Sections~\ref{sec:modeling}-\ref{sec:modeling2} and Appendix~\ref{truncated_disc}), which is emphatically disfavored by all $IRF$ fits.

However, one issue with the UV continuum arising mainly from diffuse BLR emission is that standard BLR reverberation mapping works! That is, the BLR lines lag the UV continuum, in apparent contradiction to the idea that the optical/UV continuum is produced within the BLR. For instance, \cite{BDC06} and \cite{DeR18} find a lag of $\sim 6$~days between the $H\beta$-line and the underlying continuum in NGC 4151, implying a physical separation between the optical continuum and $H\beta$ emission zones. That said, the lines and the diffuse continuum may have very different radiation patterns and the observed lags can be produced if the line emission is produced more efficiently from the illuminated front face of the BLR clouds (the magenta photons in Fig.~\ref{fig:STRUCTURE2}), while the diffuse continuum is predominantly produced on the shaded side. This weights the line response towards material on the far side of the black hole, while the continuum is weighted to the near side, with a smaller light travel time delay (red and green photons in Fig.~\ref{fig:STRUCTURE2}; see e.g. \citealt{ARR17}, comparing Figs. 6 and 7). Indeed the example of a $H\beta$ lag of $6$~days is well within the $<10$~day lag margin attributable to anisotropic clump illumination according to the simulations of Lawther et al. (2018; see their Fig.~7). On the other hand, if some lines produce a similar radiation pattern to the diffuse continuum from our perspective, we would predict a negligible lag for these lines relative to the optical; this is also supported by some observations. Taking the cases analysed by \cite{PC88} for instance, the combined HeII lag measured by Antonucci \& Cohen (1983; $-1\pm5$~days), and the MgII and CIV lags measured by Ulrich et al. (1984; $12\pm20$~days, $5\pm8$~days) are all consistent with zero, as one may expect if these lines are produced co-spatially and aligned with the diffuse continuum.

\subsection{Revisiting NGC 5548}

GD17 instead proposed that the warm Comptonisation region is vertically extended (see their Fig.~7b) so that it shields the outer disc from the hard X-ray illumination. This geometry appeared feasible as the models were based on {\tt{optxagnf}} which uses only a single warm Compton component rather than considering its radial structure as we do here. In this case our Fig.~\ref{fig:SED} shows that - while the warm Compton annuli are more luminous at smaller radii - UVW1 has equal contributions from the all radii. The inner radii could shield the outer radii from the hard X-rays, but we would still see the fast reprocessed variability from the inner warm Compton radii. Self shielding by the warm Comptonisation zone in NGC 4151 cannot hide the reprocessed signal in our model. Similar considerations show that (with hindsight) this could not work in NGC 5548 either. 

Nonetheless, in NGC 5548 the requirement for X-ray shielding was not so compelling as the X-ray bandpass only extended up to 10~keV, and was clearly affected by time variable absorption (\citealt{MKK16}). Thus it was always possible in this object that the observed X-ray variability did not track the variability of the bolometric flux, which peaks at $100$~keV, and is unaffected by obscuration. The lack of a linear $IRF$ between these signals supports this idea (see Appendix~\ref{5548}). Thus in NGC 5548 the variable UVW1 and optical emission could both be from reprocessing on a disc/warm Compton region of the (unseen) $100$~keV X-ray flux. However, the lag between the optical/UV behind the HST far-UV was also too long by a factor of 2.

GD17 explicitly tested models for the BLR contribution to the lags, and showed that this did not work to explain the UV-optical lag. However, they used a much longer lag time of 15 light days - as expected for the H$\beta$ BLR size scale given the observed luminosity - but not consistent with the actual BLR lag of a few days measured during the NGC 5548 campaign (\citealt{PFB17}, their Fig.~13; \citealt{DeR18}). They also assumed the spectrum of the BLR was dominated by the observed spectral features, rather than being a diffuse continuum from higher density clouds, and thus they underestimated the contribution of the reprocessed BLR flux. Their reverse engineered transfer functions showed that the lag of the UV and optical required material at a few light days, consistent with the observed BLR size scale in these data. However, for blackbody (rather than diffuse) emission, the thermalisation temperature is too low, which caused them to discount this possibility. 

In retrospect it seems more likely that the NGC 5548 monitoring campaign was sampling an AGN with a similar geometry to that shown in Fig.~\ref{fig:STRUCTURE2}; it was at a similarly low Eddington luminosity of $\sim0.03L_{Edd}$, and its broad H$\beta$ line is almost as weak as that in NGC 4151 \cite{DeR18}.

\section{Conclusions}
\label{sec:Conclusions}
We have built a full spectral-timing model to predict the UV lightcurve from the observed hard X-ray lightcurve in NGC 4151. The model is based on a (truncated) disc and warm Comptonisation region illuminated by the hard X-ray source. Energetically, the X-ray luminosity requires all of the gravitational energy released within the inner $90~R_g$, whereas the UV can be produced by the same accretion rate through the truncated disc/warm Compton region at larger radii, between $90$ and $390~R_g$. We use this geometric/energetic model to predict the light curve in the UVW1 band arising from illumination of the warm Compton/thin disc structure by the coronal hard X-rays. The changing hard X-ray luminosity drives variations in heating of the warm Comptonising material above the disc, as well as variations in its seed photons due to reprocessing. We show that such a geometry entirely fails to reproduce observed UVW1 variability from the observed hard X-ray ($15-50$~keV) variability. In particular, a disc geometry strongly over-predicts the coherence between these signals and under-predicts the lag. Optically thick material within a few hundred $R_g$ should respond on timescales of less than $1-2$ days and be smoothed on this timescale by the light travel time across the region. This predicts that the UV and X-ray lightcurves should be strongly correlated on timescales shorter than $1.5$~days, yet the data show only a weak correlation. The \textit{Swift} BAT band traces the bolometric luminosity of the source (unlike the `hard' X-ray band of the \textit{Swift} XRT which can be affected by the time variable absorption), and therefore the discrepancy between the model and data seen here provides the strongest confirmation yet shown that the geometry of the hard X-ray and UV sources cannot be as simple as that of a compact hot diffuse corona/lamppost illuminating optically thick material within a few hundred $R_g$. 

Instead, we find that the UV can be produced by reprocessed X-ray flux in material on size scales of the BLR. This material cannot be emitting blackbody radiation as the resulting temperatures would be too low at this size scale. However, dense clumps ($n>10^{12-15}$~cm$^{-3}$) which are optically thin rather than thick could produce a diffuse recombination continuum rather than lines or blackbody emission, as suggested by \cite{KG01} and \cite{LGK18}. Such material is seen in the line of sight as the complex, mainly neutral, X-ray absorber and is also imprinted on the optical emission lines as the H$\beta$ absorption. Following Dehghanian et al. (2019a/b) we identify this with a wind on the inner edge of the BLR. We suggest that this is also the inner edge of the optically thick accretion disc, so that inwards of this there is only the hot, optically thin gas from which the hard X-rays originate. This interpretation is supported by Chandra grating data which suggest that the `narrow' Fe~K$\alpha$ line is produced by thermal disc material extending no further in than $500~R_g$ at this luminosity (\citealt{MCZ17}), as well as by the recent work of \cite{ZMC19}, who find no evidence of relativistic Fe~K$\alpha$ line broadening associated with an inner disc. Higher spectral resolution observations with the X-ray Imaging and Spectroscopy Mission (XRISM; JAXA/NASA, probable launch 2022) and the Advanced Telescope for High-Energy Astrophysics (ATHENA; ESA, probable launch 2030) will undoubtedly reveal this structure in more detail.

\section*{Acknowledgements}
The authors thank the anonymous referee for their very helpful comments and suggestions which served to improve the manuscript. The authors thank Rick Edelson for initiating the awesome monitoring campaigns, and more specifically for the NGC 4151 light curves and spectral data. The authors also thank Bo\.{z}ena Czerny, Swayamtrupta Panda, Gary Ferland, and many attendees of the 2019 Guilin conference on Mapping the Central Regions of Active Galactic Nuclei for illuminating conversations about the BLR and reverberation mapping. RDM acknowledges the support of a Science and Technology Facilities Council (STFC) studentship through grant ST/N50404X/1. CD acknowledges the STFC through grant ST/P000541/1 for support.


\appendix

\section{Physical modeling with no soft excess region}
\label{truncated_disc}
\begin{figure*}
	\includegraphics[width=\textwidth]{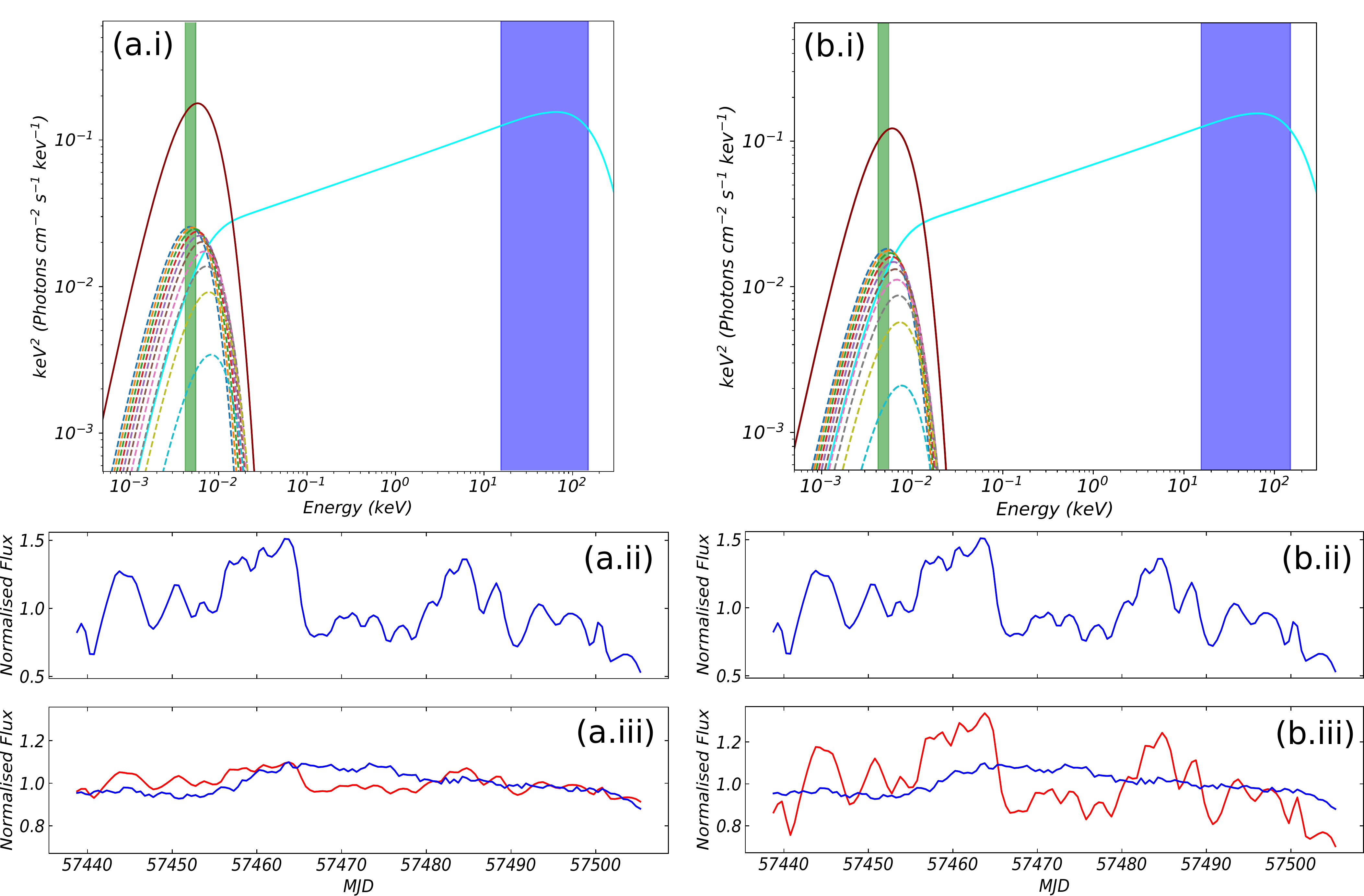}
	\caption{Panel~(a.i) shows the intrinsic SED for NGC 4151 in the case that there is no warm Compton zone, and the spectrum is only composed of a thermal disc illuminated by a central hard Compton source at a height of $h_{cor}=10$ above the black hole. Constituent thermal disc annuli are shown as dashed lines. Like the canonical model, this geometry also produces a very poor match to the true (blue) UVW1 curve in panel~(a.iii), where we see that the modeled (red) curve has a lot more of the hard Compton variability on short timescales than we find in the data. Panels~(b.i-iii) show the same model setup only now with the central source at $h_{cor}=100$, showing an even greater disparity with the UVW1 data in panel~(b.iii).}
	\label{fig:truncated_disc_all4}
\end{figure*}

In this Appendix we highlight that the modeled UVW1 curve we recovered in Section~\ref{sec:modeling2} is insensitive to the form of accretion disc assumed at small radii, such that even if the disc within $1$~light day does not have a warm Compton atmosphere, the fact that it must dominate the UV band means that the resultant UVW1 curve is still strongly dissimilar to the data.

If our data have undergone more complex reddening than we anticipated in Section~\ref{sec:modeling}, it becomes possible that there is no warm Compton component at all, and the intrinsic SED is actually dominated by only a thermal Shakura-Sunyaev disc and a hard Compton component. By including a second, multiplicative {\tt{zredden}} component in addition to the other components listed in Section~\ref{sec:modeling_spectral}, and tying $r_{warm}$ to $r_{hot}$, we find that we can in fact fit the spectrum with the same quality as the canonical model, now omitting the warm Compton zone. We show the intrinsic SED (not including the complex absorption, host galaxy dilution and reddening) in Fig.~\ref{fig:truncated_disc_all4}, panel~(a.i), where we have fixed the coronal height to $h_{cor}=10$, and the fit inner disc radius is $60~R_g$. We see in panel~(a.iii) that the predicted light curve from this geometry is nothing like that which is observed. The disparity becomes far more pronounced when the coronal height is increased to $h_{cor}=100$ in panels~(b.i-iii), where the variability amplitude as well as shape is now entirely wrong.

Combined with the results of Sections~\ref{sec:modeling}~and~\ref{sec:modeling2}, we conclude with a high degree of certainty that, as long as optically thick material at sub-light-day size scales dominates the UV band - whether that material is thermal or weakly Comptonised - these geometries will fail to correctly predict the UV light curve, and so should be ruled out.

\section{Impulse Response Recovery in NGC 5548}
\label{5548}
\begin{figure*}
	\includegraphics[width=\textwidth]{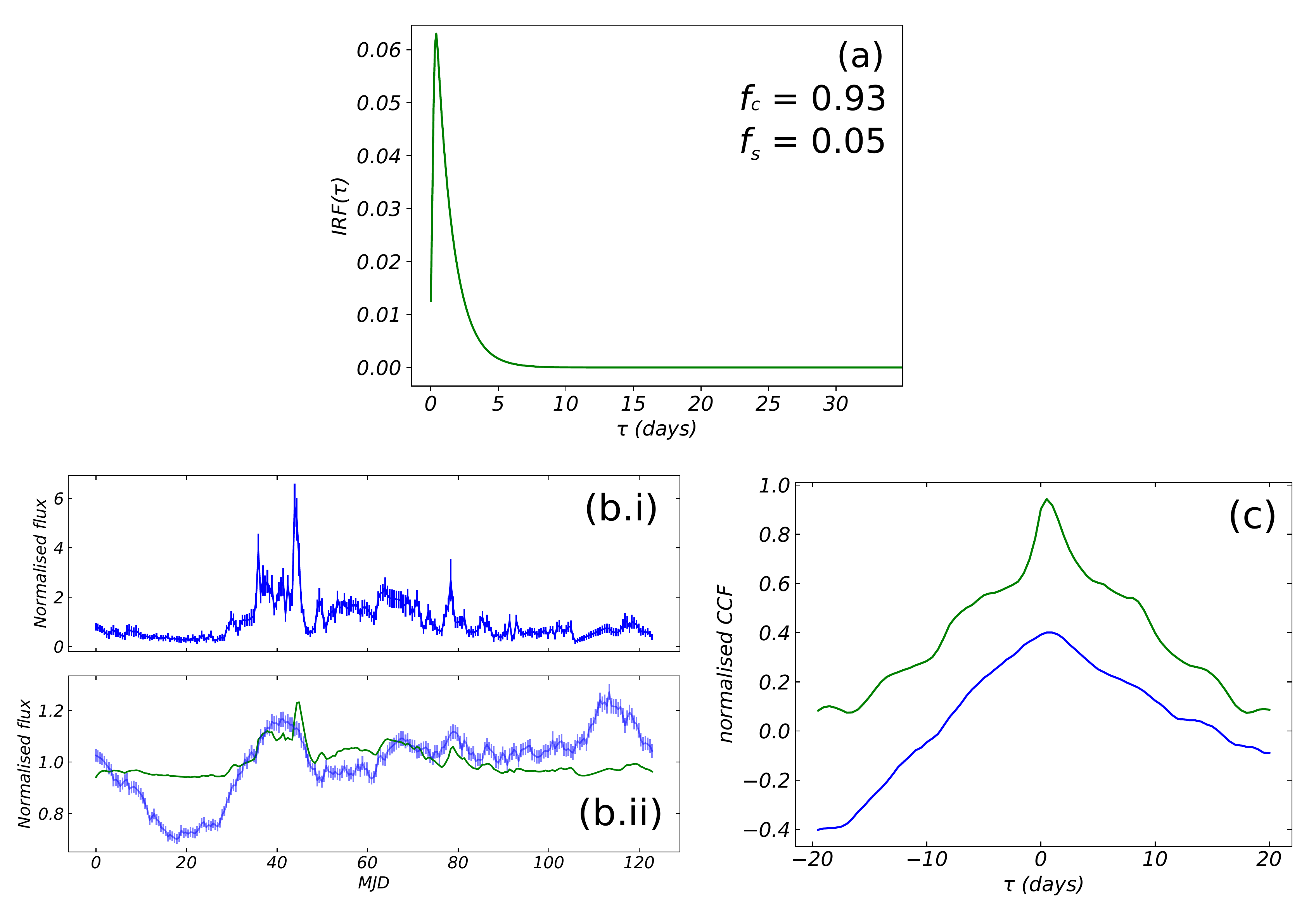}
	\caption{Modeled UVW1 light curve using phenomenological impulse responses to the \textit{Swift} XRT X3 curve for the NGC 5548 data of Gardner \& Done (2017; from \citealt{E15}). Panel~(a): Impulse response function used to fit the UVW1 light curve, with functional form in equation~(\ref{eq:fiducialIRF2}). Panel~(b.i): Interpolated \textit{Swift} XRT X3 light curve. Panel~(b.ii): Interpolated observed UVW1 light curve denoted in blue. Modeled UVW1 light curves denoted in green, where the UV emitting component(s) is directly modulated by the \textit{Swift} XRT X3 lightcurve convolved with the $IRF$ of panel~(a). Panel~(c): Cross correlation functions between \textit{Swift} XRT X3 and UVW1 light curves. The observed CCF is denoted in blue. The modeled CCF resulting from the $IRF$ in Panel~(a) is denoted in green.}
	\label{fig:reverseengineered5548}
\end{figure*}

Here we apply the impulse response recovery method of Section~\ref{reverseengineering} to the 2014 UVW1 and \textit{Swift} XRT light curves of NGC 5548 (\citealt{E15}), originally investigated with physical modeling efforts in GD17. In that work it was found that the sharply peaked variations in the $0.8-10$~keV band and lack thereof in the UVW1 band resulted in the simple disc reverberation model failing to replicate the UVW1 curve from the hard X-ray input. We therefore test whether the phenomenological model used here can recover the UVW1 curve (and associated impulse response) when the driving signal is associated with the $0.8-10$~keV band in NGC 5548.

We assume integrated hard Compton-to-disc contributions in the UVW1 band using the SED of Fig.~3(a) of GD17, and that the $0.8-10$~keV band is contributed to exclusively by the hard Compton component. Equation~(\ref{FUVW1}) therefore applies. We also prescribe the exponentially modified double-Gaussian functional form for the impulse response for maximum fit freedom, so that equation~(\ref{eq:fiducialIRF2}) applies.

We find that this technique is unable to reproduce the UVW1 light curve from the hard X-ray input, as shown by the highly contrasting UVW1 light curves in Fig.~\ref{fig:reverseengineered5548}(b.ii). No matter the shape of the impulse response function, the highly peaked hard X-ray light curve cannot be smoothed sufficiently to reproduce the UVW1 light curve without inducing a strong lag, which the observed light curves do not appear to exhibit. We therefore end up with a model UVW1 curve which exhibits only a short lag as required, but with insufficient smoothing. Furthermore, other features in the observed UVW1 curve (e.g. the peak at 115 days) have absolutely no counterpart in the hard X-ray light curve, adding to the weight of evidence which suggests the sources of emission in these bands do not continuously communicate in NGC 5548, seemingly unlike in NGC 4151.

\bsp
\label{lastpage}

\begin{thebibliography}{99}


\bibitem[\protect\citeauthoryear{Almeyda et al.}{2017}]{ARR17}
Almeyda T., Robinson A.,Richmond M., Vazquez B., Nikutta R., 2017, ApJ, 841, 1


\bibitem[\protect\citeauthoryear{Antonucci \& Cohen}{1983}]{AC83}
Antonucci R.R.J. \& Cohen R.D., 1983, ApJ, 271, 564

\bibitem[\protect\citeauthoryear{Ar\'{e}valo et al.}{2009}]{A09}
Ar\'{e}valo P., Uttley P., Lira P., Breedt E., McHardy I.M., Churazov E., 2009, MNRAS, 397, 2004

\bibitem[\protect\citeauthoryear{Arnaud, Borkowski \& Harrington}{1996}]{ABH96}
Arnaud K., Borkowski K.J., Harrington J.P., 1996, ApJ, 462, L75

\bibitem[\protect\citeauthoryear{Arnaud \& Raymond}{1992}]{AR92}
Arnaud M., Raymond J., 1992, ApJ, 398, 394





\bibitem[\protect\citeauthoryear{Baron et al.}{2016}]{BSP16}
Baron D., Stern J., Poznanski D., Netzer H., 2016, ApJ, 832 (1), 8

\bibitem[\protect\citeauthoryear{Barvainis}{1987}]{B87}
Barvainis R., 1987, ApJ, 320, 537



\bibitem[\protect\citeauthoryear{Beuchert et al.}{2017}]{BMD17}
Beuchert T., Markowitz A.G., Dauser T. et al., 2017, A\&A, 603, A50




\bibitem[\protect\citeauthoryear{Bentz et al.}{2006}]{BDC06}
Bentz M.C., Denney K.D., Cackett E.M. et al., 2006, ApJ, 651, 775

\bibitem[\protect\citeauthoryear{Bentz et al.}{2013}]{BDG13}
Bentz M.C., Denney K.D.,Grier C.J. et al., 2013, ApJ, 767 (2), 149



\bibitem[\protect\citeauthoryear{Boissay, Ricci \& Paltani}{2016}]{BRP16}
Boissay R., Ricci C. \& Paltani S., 2016, A\&A, 588, A70

\bibitem[\protect\citeauthoryear{Buisson et al.}{2018}]{BLA18}
Buisson D.J.K., Lohfink A.M., Alston W.N. et al., 2018, MNRAS, 475 (2), 2306




\bibitem[\protect\citeauthoryear{Chelouche, Nu\~{n}ez \& Kaspi}{2018}]{CNK18}
Chelouche D., Nu\~{n}ez F.P., Kaspi, S., 2018, \textit{Nat. Astron.}




\bibitem[\protect\citeauthoryear{Davis, Woo \& Blaes}{2007}]{DWB07}
Davis S.W., Woo J.-H., Blaes O.M, 2007, ApJ, 668 (2), 682


\bibitem[\protect\citeauthoryear{De Rosa et al.}{2018}]{DeR18}
De Rosa G., Fausnaugh M.M., Grier C.J. et al., 2018, ApJ, 866, 133

\bibitem[\protect\citeauthoryear{Dehghanian et al.}{2019a}]{DFK19}
Dehghanian M., Ferland G.J., Kriss G.A. et al., 2019, ApJ, 877 (2), 119

\bibitem[\protect\citeauthoryear{Dehghanian et al.}{2019b}]{DFP19}
Dehghanian M., Ferland G.J., Peterson B.M. et al., 2019, ApJL, 882, 2





\bibitem[\protect\citeauthoryear{Done, Gierli\'{n}ski \& Kubota}{2007}]{DGK07}
Done C., Gierli\'{n}ski M., Kubota A., 2007, A\&ARv, 15, 1


\bibitem[\protect\citeauthoryear{Done et al.}{2012}]{D12}
Done C., Davis S.W., Jin C., Blaes O., Ward M., 2012, MNRAS, 420, 1848


\bibitem[\protect\citeauthoryear{Edelson et al.}{2015}]{E15}
Edelson R., Gelbord K., Horne K. et al., ApJ, 806, 129

\bibitem[\protect\citeauthoryear{Edelson et al.}{2017}]{E17}
Edelson R., Gelbord K., Cackett E. et al., ApJ, 840, 41 (E17)


\bibitem[\protect\citeauthoryear{Ezhikode et al.}{2018}]{EGD17}
Ezhikode S.H., Gandhi P., Done C. et al., 2018, MNRAS, 472 (3), 3492

\bibitem[\protect\citeauthoryear{Fabian et al.}{1989}]{F89}
Fabian A.C., Rees M.J., Stella L., White N.E., 1989, MNRAS, 238, 729




\bibitem[\protect\citeauthoryear{Frank, King \& Raine}{2002}]{FKR}
Frank J., King A., Raine D., 2002, \textit{Accretion Power in Astrophysics: Third Edition}, Cambridge University Press, pp.~83-84


\bibitem[\protect\citeauthoryear{Galeev, Rosner \& Vaiana}{1979}]{GRV79}
Galeev A., Rosner R., Vaiana G., 1979, ApJ, 229, 318


\bibitem[\protect\citeauthoryear{Garc{\'\i}a et al.}{2019}]{GKW19}
Garc{\'\i}a J.A., Kara E., Walton D. et al., 2019, ApJ, 871, 88


\bibitem[\protect\citeauthoryear{Gardner \& Done}{2017}]{GD17}
Gardner E., Done C., 2017, MNRAS, 470, 3591 (GD17)





\bibitem[\protect\citeauthoryear{Gierli{\'n}ski \& Done}{2004}]{GD04b}
Gierli{\'n}ski M., Done C., 2004, MNRAS, 349, L7






\bibitem[\protect\citeauthoryear{Haardt \& Maraschi}{1993}]{HM93}
Haardt F., Maraschi L., 1993, ApJ, 413, 507












\bibitem[\protect\citeauthoryear{Inoue \& Doi}{2018}]{ID18}
Inoue Y., Doi A., 2018, ApJ, 869 (2), 114

\bibitem[\protect\citeauthoryear{Jin et al.}{2012}]{JWD12}
Jin C., Ward M., Done C., Gelbord J., 2012, MNRAS 420, 1825

\bibitem[\protect\citeauthoryear{Kaiser et al.}{2000}]{KBH00}
Kaiser M.E., Bradley L.D., Hutchings J.B.  et al., 2000, ApJ, 528, 260



\bibitem[\protect\citeauthoryear{Keck et al.}{2015}]{KBB15}
Keck M.L., Brenneman L.W., Ballantyne D.R. et al., 2015, ApJ, 806, 149



\bibitem[\protect\citeauthoryear{Kishimoto et al.}{2013}]{KHA13}
Kishimoto M., H\"{o}nig S.F., Antonucci R. et al., ApJ \textit{Letts.}, 2013, 775 (2), L36

\bibitem[\protect\citeauthoryear{Korista \& Goad}{2001}]{KG01}
Korista K.T., Goad M.R., 2001, ApJ, 553, 695





\bibitem[\protect\citeauthoryear{Kubota \& Done}{2018}]{KD18} (KD18)
Kubota A., Done C., 2018, MNRAS, 480, 1247


\bibitem[\protect\citeauthoryear{Laor \& Davis}{2014}]{LD14}
Laor A., Davis S.W., 2014, MNRAS, 438 (4), 3042

\bibitem[\protect\citeauthoryear{Lawrence}{2018}]{L18}
Lawrence A., 2018, Nature Astronomy, 2, 102-103

\bibitem[\protect\citeauthoryear{Lawther et al.}{2018}]{LGK18}
Lawther D., Goad M.R., Korista K.T., Ulrich O., Vestergaard M., 2018, MNRAS, 481 (1), 533


\bibitem[\protect\citeauthoryear{Lubi\'{n}ski et al.}{2002}]{LZW10}
Lubi\'{n}ski P., Zdziarski A.A., Walter R., Paltani S., Beckmann V., Soldi S., Ferrigno C., Courvoisier T.J.-L., 2010, MNRAS, 408 (3), 1851





\bibitem[\protect\citeauthoryear{Magdziarz et al.}{1998}]{M98}
Magdziarz P., Blaes O.M., Zdziarski A.A., Johnson W.N., Smith D.A., 1998, MNRAS, 301, 179







\bibitem[\protect\citeauthoryear{Matt et al.}{2014}]{M14}
Matt G., Marinucci A., Guainazzi M. et al., 2014, MNRAS, 439, 3016


\bibitem[\protect\citeauthoryear{Mehdipour et al.}{2011}]{MBR11}
Mehdipour M., Branduardi-Raymont G., Kaastra J.S. et al., 2011, A\&A, 534, A39

\bibitem[\protect\citeauthoryear{Mehdipour et al.}{2015}]{MKK15}
Mehdipour M., Kaastra J.S., Kriss G.A. et al., 2015, A\&A, 575, A22

\bibitem[\protect\citeauthoryear{Mehdipour et al.}{2015}]{MKK16}
Mehdipour M., Kaastra J.S., Kriss G.A. et al., 2016, A\&A, 588, A139

\bibitem[\protect\citeauthoryear{Mewe, Gronenschild \& van den Oord}{1985}]{MGvdO85}
Mewe R., Gronenschild E.H.B.M., van den Oord, G.H.J., 1985, A\&A Suppl., 62, 197


\bibitem[\protect\citeauthoryear{Miller et al.}{2017}]{MCZ17}
Miller J.M., Cackett E., Zoghbi A. et al., 2017, ApJ, 865 (2), 97






\bibitem[\protect\citeauthoryear{Nandra et al.}{2007}]{N07}
Nandra K., O'Neill P.M., George I.M., Reeves J.N., 2007, MNRAS, 382, 194





\bibitem[\protect\citeauthoryear{Noda \& Done}{2018}]{ND18}
Noda H., Done C., 2018, MNRAS, 480, 3898




\bibitem[\protect\citeauthoryear{Novikov \& Thorne}{1973}]{NT73}
Novikov I.D., Thorne K.S., 1973, blho.conf, 343

\bibitem[\protect\citeauthoryear{Oknyansky et al.}{2014}]{OMT14}
Oknyansky V.L., Metlova N.V., Taranova O.G., Shenavrin V.I., Artamonov B.P., Gaskell C.M., Guo Di-Fu, 2014, Odessa Astronomical Publications, 27, 47



\bibitem[\protect\citeauthoryear{Pei et al.}{2017}]{PFB17}
Pei L., Fausnaugh M.M., Barth A.J. et al., 2017, ApJ, 837, 131

\bibitem[\protect\citeauthoryear{Penston \& Perez}{1984}]{PP84}
Penston M.V., Perez E., 1984, MNRAS, 211, 33


\bibitem[\protect\citeauthoryear{Petrucci et al.}{2013}]{POP13}
Petrucci P.-O., Paltani S., Malzac J. et al., 2013, A\&A, 549, A73

\bibitem[\protect\citeauthoryear{Petrucci et al.}{2018}]{POP18}
Petrucci P.O., Ursini F., De Rosa A., Bianchi S., Cappi M., Matt G., Dadina M., Malzac J., 2018, A\&A, 611, A59

\bibitem[\protect\citeauthoryear{Peterson \& Cota}{1988}]{PC88}
Peterson B.M., Cota S.A., 1988, ApJ, 330, 111

\bibitem[\protect\citeauthoryear{Peterson \& Horne}{2004}]{PH04}
Peterson B.M., Horne K., 2004, Astronomische Nachrichten, 325, 248


\bibitem[\protect\citeauthoryear{Porquet et al.}{2004}]{PRO04}
Porquet D., Reeves J.N., O'Brien P., Brinkmann W., 2004, A\&A, 422, 85

\bibitem[\protect\citeauthoryear{Porquet et al.}{2018}]{PRM18}
Porquet D., Reeves J.N., Matt G. et al., 2018, A\&A, 609, A42














\bibitem[\protect\citeauthoryear{Richards et al.}{2003}]{RHV03}
Richards G.T., Hall P.B., Vanden Berk D.E. et al., 2003, ApJ, 126 (3), 1131

\bibitem[\protect\citeauthoryear{Ruan et al.}{2019}]{RAE19}
Ruan J.J., Anderson S.F., Eracleous M. et al., 2019, submitted, arXiv:1903.02553


\bibitem[\protect\citeauthoryear{Shakura \& Sunyaev}{1973}]{SS73}
Shakura N.I., Sunyaev R.A., 1973, A\&A, 24, 337

\bibitem[\protect\citeauthoryear{Shapovalova et al.}{2008}]{SPC08}
Shapovalova A.I., Popovi{\'c} L.C., Collin S. et al., 2008, A\&A, 486, 99



\bibitem[\protect\citeauthoryear{Shu, Yaqoob \& Wang}{2010}]{SYW10}
Shu X.W., Yaqoob T., Wang J.X., 2010, ApJ Suppl., 187, 581

\bibitem[\protect\citeauthoryear{Shull, Stevans \& Danforth}{2012}]{SSD12}
Shull M.J., Stevans M., Danforth C.W., 2012, ApJ, 752 (2), 162




\bibitem[\protect\citeauthoryear{Telfer et al.}{2002}]{TZK02}
Telfer R.C., Zheng W., Kriss G.A., Davidsen A.F., 2002, ApJ, 565 (2), 772





\bibitem[\protect\citeauthoryear{ULrich et al.}{1984}]{UBB84}
Ulrich M.H., Boksenber A., Bromage G.E. et al., 1984, MNRAS, 206 221







\bibitem[\protect\citeauthoryear{Veledina}{2016}]{V16}
Veledina A., 2016, ApJ, 832, 181



\bibitem[\protect\citeauthoryear{Welsh \& Horne}{1991}]{WH91}
Welsh W.F., Horne K., 2016, ApJ, 379, 586





\bibitem[\protect\citeauthoryear{Yuan, Quataert \& Narayan}{2003}]{YQN03}
Yuan F., Quataert E. \& Narayan R., 2003, ApJ, 598 (1), 301



\bibitem[\protect\citeauthoryear{Zheng et al.}{1997}]{ZKT97}
Zheng W., Kriss G.A., Telfer R.C, Grimes J.P., Davidsen A.F., 1997, ApJ, 475 (2), 469

\bibitem[\protect\citeauthoryear{Zoghbi, Miller \& Cackett}{2019}]{ZMC19}
Zoghbi A., Miller J., Cackett E., 2019, ApJ, 884, 26

\bibitem[\protect\citeauthoryear{Zy{\.c}ki, Done \& Smith}{1999}]{ZDS99}
Zy{\.c}ki P.T., Done C., Smith D.A., 1999, MNRAS, 305, 231

\end{thebibliography}
\end{document}